\documentstyle[prb,aps,tighten,draft,epsf]{revtex}

\newcommand{\fig}[2]{\epsfxsize=#1\epsfbox{#2}}
 \newcommand{\beq}{\begin{equation}}
 \newcommand{\eeq}{\end{equation}}
 \newcommand{\beqn}{\begin{eqnarray}}
 \newcommand{\eeqn}{\end{eqnarray}}
 \newcommand{\beqa}{\begin{eqnarray}}
 \newcommand{\eeqa}{\end{eqnarray}}
 \newcommand{\bi}{\bibitem}
 \newcommand{\n}{\newline}
 \newcommand{\nn}{\nonumber\\}
 \newcommand{\bc}{\begin{center}}
 \newcommand{\ec}{\end{center}}

\begin{document}
\author{Laurent Laloux$^*$ and Pierre Le Doussal$^{**}$}
\address{{$^*$} Laboratoire de Physique Th\'{e}orique et Hautes
Energies,\cite{auth2}
4 Place Jussieu, 75252 Paris Cedex 05, France}
\address{$^{**}$ CNRS-Laboratoire de Physique Th\'{e}orique de l'Ecole Normale
 Sup\'{e}rieure \cite{cnrs}, 24 rue Lhomond 75231 Paris France}

\title{Aging and diffusion in low dimensional environments}
\date{\today}
\maketitle

\begin{abstract}
 We study out of equilibrium dynamics and aging for
 a particle diffusing in one dimensional environments,
 such as the random force Sinai model, as a toy model
 for low dimensional systems.  We study fluctuations
 of two times $(t_w,  t)$ quantities from the probability
 distribution $Q(z,t,t_w)$ of the relative displacement
 $z = x(t) - x(t_w)$ in the limit of large waiting time
 $t_w \to \infty$ using numerical and analytical techniques. 
 We find three generic large time regimes:
 (i) a quasi-equilibrium regime (finite $\tau=t-t_w$)
 where $Q(z,\tau)$ satisfies a general FDT equation
 (ii) an asymptotic diffusion regime for large time separation
 where $Q(z) dz \sim \overline{Q}[L(t)/L(t_w)] dz/L(t)$
 (iii) an intermediate ``aging'' regime for intermediate
 time separation ($h(t)/h(t_w)$ finite), with 
$Q(z,t,t') = f(z,h(t)/h(t')) $. 
 In the 
 unbiased Sinai model we find numerical evidence
 for regime (i) and (ii), and for (iii) with 
 $\overline{Q(z,t,t')} = Q_0(z) f(h(t)/h(t'))$ and $h(t) \sim \ln t$.
 Since $h(t) \sim L(t)$ in Sinai's model there is
 a singularity in the diffusion regime to allow for
 regime (iii). A directed model, related to the biased 
 Sinai model is solved and shows (ii) and (iii) with
 strong non self-averaging properties. Similarities and
 differences with mean field results are discussed.
 A general approach using scaling of next highest encountered
 barriers is proposed to predict aging
 properties, $h(t)$ and $f(x)$ in landscapes
 with fast growing barriers. It accounts qualitatively for
aging in Sinai's model. 
 We also identify a mecanism for aging in low dimensional phase space
 corresponding to an almost degeneracy of barriers.
 We illustrate this mechanism by introducing a new exactly solvable
 model, with barriers and wells, 
 which shows clearly diffusion and aging regimes with a rich variety
 of functions $h(t)$.
 \end{abstract}
 
 \widetext

 \section{Introduction}

 There is presently considerable interest in 
 out of equilibrium dynamical processes. For systems without 
 quenched disorder these are important to understand phenomena
 such as coarsening and domain growth \cite{bray}.
In systems such as spin glasses
\cite{Vihaoc,ocio,Ri,Ri2,huse_fisher}, random fields, interfaces,
glasses in vortex systems \cite{revue_russes,Legi3,Legi},
which are dominated by 
quenched disorder and ultra slow relaxations, a detailed understanding of 
 out of equilibrium dynamics becomes absolutely necessary to make contact with
 numerical simulations and experiments.
These usually involve studying relaxation dynamics from an initial 
configuration at $t=0$ (e.g uncorrelated) and asking about
 correlations in the systems between two later times $t'$ 
 (also called $t_w$ the waiting time) and $t < t'$.
An interesting question to ask is what happens when 
 both $t$ and $t'$ are taken to infinity. Since there are of course
 many ways to take $t$, $t'$ to infinity and one wishes to
 classify the possible regimes.
 
 In out of equilibrium situations the usual properties
 of equilibrium dynamics do not hold.  Such properties are
 the time translational invariance (TTI) , i.e the dependence
 on $t-t'$ only of the correlation functions
 as well as the fluctuation dissipation theorem (FDT)
 which relates linear response to time derivatives of
 correlations functions.  A first question is then to ask
how to take $t$, $t'$ to infinity and still
recover an equilibrium regime.
 
 Glassy systems with quenched disorder were found to 
 exhibit a variety of non equilibrium properties often
 generically termed ``aging''. Loosely it means that the
 properties of the system are governed by the age 
 of the system $t_w$, i.e the time after the quench
\cite{aging,Go,Dofeio}
 For instance it is expected that correlation functions in these systems
 have dependences such has $t/t_w$.  This type of dependence
 is also found in simpler out of equilibrium systems such as coarsening in
 spin systems without disorder \cite{bray} which also exhibit dependences 
 of correlation functions of the form $L(t)/L(t_w)$.
 This dependence originates from the growth of domains 
 of size $L(t)$ and diffusion of the domain walls. A stronger
 form of aging seems to be observed \cite{Vihaoc,ocio}, and was proposed
 for spin glasses where the linear response shows memory effects, 
 e.g the remanent linear magnetization 
 after applying a field during time $t_w$ decays
 very slowly over a time scale set only by $t_w$.
Other puzzling phenomena such as memory under thermal cycling are observed
\cite{Vihaoc,Ri,Ri2}.
 There is at present no theory which would account fully for
all these phenomena \cite{sibani}, and understanding is only partial.
 Ideas and scaling arguments borrowed from domain growth and 
 coarsening were also applied to disordered models (``droplet
 picture'' \cite{huse_fisher}),
but it is unclear whether they can account for all situations.
More recently exact solutions were obtained \cite{Cuku}
for the out of equilibrium
 dynamics of several mean field models \cite{Cule,Cukule}
, some of which where found to exhibit
 strong aging properties.  It makes several
 non trivial predictions \cite{Cuku} for both the correlation function $C(t,t')$ and
 the response function $R(t,t')$ of these
 mean-field models {\it in the limit of large times}
 $t$,$t'$. The asymptotic time regime in these
 models has been successfully resolved, under some
 physical assumptions, and turns out to be
 in direct correspondence with Parisi's static replica
 symmetry breaking solution (with some important differences).
However the matching of the
 small time regime to the asymptotic one remains
 problematic. A general picture for aging dynamics in 
 mean field was proposed \cite{Cuku}.  The resemblance with some of
 the features observed in spin glass experiments seems encouraging,
 though many points remain unclear
\cite{leto_experiments}. 
 There has been some attempts to 
 classify the various aging behaviours, and to differentiate the
 domain growth coarsening type of ``aging'' from a stronger
 type analogous to what is found in mean field models
\cite{barrat}.
 Despite these recent advances, there is
 however at present no detailed microscopic understanding of
 the aging phenomena.
 
 An elegant microscopic mechanism for aging was proposed
 some time ago by Feigelman and Vinokur (FV) \cite{feigelman_aging_1d}
in the context
 of diffusion in a one dimensional environment. Using a semi-quantitative analysis,
 they proposed that traps with wide distribution of waiting times (a diverging first
 moment $<\tau> =\infty$ ) would naturally lead to aging phenomena and waiting time dependence.
 The idea is that because $<\tau> =\infty$,
 at time $t_w$ the system is typically in a trap of release
 time $\sim t_w$ and thus the diffusing particle
 sees potential barriers which effectively grows with $t_w$.
 The Feigelman-Vinokur trap model was later 
 popularized by Bouchaud \cite{Bo} and
 generalized, on the basis of previous work on
 wide distributions of waiting times
\cite{result_1d_epl,bouchaud_1d,bouchaud_correlated,pld_thesis,machta_diffusion}
 to describe various situations, some inspired by the statics of mean field
models \cite{dean}. 
 It is important to note, however, that 
 in the one dimensional Sinai model with a bias
 originally studied by FV and in Ref. \cite{bouchaud_1d}
the wide distributions can be shown to
 be {\it dynamically} generated and not artificially put by hand. 
 Despite the phenomenological appeal of the FV
 mecanism for aging it is unclear how far it can be pushed \cite{Bo}
to describe all the physics of glasses and provide a non artificial, dynamically generated mecanism for aging.
 In fact rather different scenarios were proposed to understand more microscopically
 mean field dynamics \cite{laloux}
 
 In this paper we study some simple one dimensional diffusion model,
 such as Sinai's model and investigate in some details their out of equilibrium
 dynamics. We will not attempt to propose any phenomenological model for aging
in real glasses but simply study the extreme case of low dimensional
 phase space and identify the various possible large time regimes. 
These low dimensional models are dominated by activated dynamics over 
energy barriers and
 stand at the opposite end from mean field models.
 One can hope that they serve as toy models for diffusion in 
 low dimensional space and give some insight into finite dimension.
In particular, one can check whether some of the ideas 
 introduced in mean-field theory carry through. Specifically, 
mean field generalizations of Sinai's model (i.e diffusion in
a $d$ dimensional random potential with $d \to \infty$) were solved
\cite{Cule}
and can be directly compared. The advantage
 of this one dimensional model is that the correlation 
 and response functions can in principle be 
 computed numerically up to very large times. Note that a simulation 
of Sinai's model was performed recently in \cite{toyaging}, but with a much too slow algorithm
to reach significant times, as we will discuss. A
first summary of the present study was contained in \cite{laloux_phd}.

It becomes clear in our study that here the {\it full probability
 distribution} should be studied, while in mean field theory it is enough
to study the second moment. 
For a diffusing particle the interesting quantity 
is the distribution $Q(z,t,t')$ of displacements $z=x(t) - x(t')$
 between times $t$ and $t'$ (the particle having
 started at $t=0$ at a random position)
 with a translational or configurational average
 of $Q$. The second moment $B(t,t') = <z^2>_Q$
 was computed in mean field \cite{Cule}. The results are as
follows for $t \to \infty$ and $t'\to \infty$. There is:

(i) a {\it quasi equilibrium regime} for finite $\tau=t-t'$
where TTI ($B(t,t')= B(\tau)$) and FDT theorems hold.
In that regime displacements are bounded since $B(\tau \to \infty) = b_0$. 

(ii) for more separated times $B(t,t')>b_0$ keeps growing. This is the 
{\it aging regime} where $B(t,t')$ remains a fixed number $B(t,t')=B$
provided $t$ and $t'$ grow in some well defined way,
i.e with $h(t)/h(t')$ a fixed number (a function of $B$).
In that very non trivial regime some new generalized FDT theorems
hold, according to the general theory of \cite{Cuku}.
The function $h(t)$ is not determined by the large time mean field 
solution, and at present must be determined numerically.
The number $B$ can eventually be chosen as large
as wanted. For large $B$ this regimes crosses over into a diffusion 
regime. Interestingly there is a singularity at the beginning
of the aging regime with a non trivial exponent $\beta$, i.e the 
function $B(t,t')=B[h(t)/h(t')]$ is non analytic as a function of
$h(t)/h(t') -1$. A singularity at the beginning of the aging regime
is indeed found in a variety of experimental glassy systems.
The exponent $\beta$, relevant for mode coupling
theories of real glasses \cite{Go}, was obtained analytically in
\cite{Cule}.

Note also that in mean field there is some mathematical correspondance between two times
$t,t'$ and replica pairs $a,b$. There, roughly speaking the quasi equilibrium regime (i)
was found to correspond to the replica symmetric part of the static
solution, while the aging regime (ii) corresponded to the
RSB part of the static solution. The correspondence is in fact not
perfect because dynamical quantities do not always coincide with
their static counterpart \cite{Cuku,Kiho1,Cule}.

It is important
to know what remains of the above dynamical mean field scenario in low dimensional models,
such as the models studied here, and we will attempt to give some elements of answer.
Obviously this scenario will be modified since we now have to deal with full
distributions $Q(z,t,t')$. One would like e.g. to identify possible mechanisms
which can reproduce aging with non trivial functions $h(t)$ and 
exponents $\beta$.

We will use Sinai's model as a starting point.
It is interesting because it is related to coarsening models
with disorder. For instance it can model a single interface in a random 
field model, or motion of kinks in vortex lines, or dislocation
loops, in presence of point disorder. The dynamics of Sinai's
 model has been studied extensively 
 \cite{derrida,solomon,derrida_pomeau,bouchaud_1d,sinai}.
 The model was shown to
 exhibit ultra-slow diffusion $\overline{<x(t)^2>} \sim (\log t)^4$.
 The response to a driving force $f$ was also shown to be quite
 anomalous, with several phases. There is a threshold force
 such that for $f<f_c$ the velocity vanishes $V=0$ and
 diffusion is sublinear $\overline{<x(t)^2>} \sim t^{2 \mu}$
 with a continuously variable exponent $\mu=f/f_c$. For 
 $f>f_c$ one has $V \sim (f-f_c)$ and similar transitions
 in higher order moments of the displacement take place at
 larger $f$. The physics of this anomalous response 
 was also understood and shown to be related
\cite{bouchaud_1d,sinai} to the existence of barriers with
 an exponential distribution of energy heights, resulting in
 power-law distributions of trapping times and 
 Levy distributions for first passage times.

Let us illustrate
some of the questions by some simple consideration.
In the case of  the case
of an applied bias, even the {\it first moment} of the
 displacement already contains interesting information.
 It is known \cite{solomon,bouchaud_1d,sinai} that one has exactly, at large times
 $<\overline{x(t) - x(0)>} = C t^{\mu}$
 where $C$ depends on the details of the model. The aging nature
of this expression is clear since one can write:
 \beq
 <\overline{x(t) - x(t')>} = C ( t^{\mu} - t'^{\mu} ) 
 \sim C \mu \frac{\tau}{t_w^{1-\mu}} 
 \eeq
 for $t'=t_w$, $t=t_w + \tau$. This indicates that there is an aging
 regime in this model since one can take $\tau \to \infty$
 and $t_w \to \infty$ keeping $<\overline{x(t) - x(t')>}$ 
 a fixed number, provided:
 \beq
 \tau \sim t_w^{1-\mu}
 \eeq
 In particular, in the limit where $t_w$ 
 goes to infinity first, there is no motion at all 
 in any finite time interval! We also note that the above result
can be rewritten as:
 \beq  \label{form}
 \overline{<x(t) - x(t')>} = C \ln \frac{h(t)}{h(t')}
 \eeq
 where $h(t)=\exp(t^{\mu})$. This form (\ref{form}) (with $h(t)$ unspecified)
is of the type found in large time solutions of mean field
 models as mentionned above. Similarly, subaging forms for
$h(t)=\exp((\ln t)^a)$ with $a>1$, 
as observed in experiments \cite{leto_experiments}, could
in principle originate from logarithmic diffusion process. Indeed one can write:

\beq
\ln^a t - \ln^a t' = \ln \frac{h(t)}{h(t')}  \sim \frac{ \tau }{ (t_w/\ln^{a-1} t_w) }
\eeq

with the observed $h(t)$. One would thus like to investigate these
analogies further.

We will thus first study Sinai's model in Section \ref{sinaisec}
(symmetric model) and in Section \ref{directed} (directed model).
Then we will give in  Section \ref{barriers} a general discussion
of barrier mechanisms for aging which we believe allow to understand
the results obtained on Sinai's model. We introduce a method
to estimate aging functions by looking at the sequences of next highest
barriers. Finally, in Section \ref{solvable} we introduce and
solve a model based on these considerations 
which exhibits non trivial againg and diffusion behaviour.

\section{Symmetric Sinai model} \label{sinaisec}

\subsection{the model}
 
In this Section we study the one dimensional Sinai model
\cite{sinai,derrida,solomon,derrida_pomeau,golosov,bouchaud_1d}. In its continuous
version it is described by the Langevin equation:

\begin{eqnarray}
\frac{dx}{dt}= F(x(t)) + \eta(t) 
\end{eqnarray}

where $<\eta(t) \eta(t')> = 2T \delta(t-t')$ is the thermal noise 
and $\overline{F(x) F(x')} = \sigma \delta(x-x')$ is a quenched
random force which is gaussian and uncorrelated. It may have an average
$\overline{F(x)}=f$. Writing 
$F(x) = - dU(x)/dx$ it describes
the thermal motion of a particle in a one dimensional
gaussian random potential landscape $U(x)$. The landscape
$(x,U(x))$ can be seen itself as a trajectory of a random walker
($x$ playing then the role of the time). In the general case
$f>0$ the lanscape is tilted (the walker experiences a bias) but
we will only consider here the case $f=0$ (symmetric Sinai model).
Then the random potential has long range correlations:

\begin{eqnarray}
\overline{(U(x)-U(y))^2} = \sigma |x-y|
\end{eqnarray}

Thus barriers grow with scale which
results in an anomalously slow
dynamics $<(x(t)-x(0))^2> \sim (\ln t)^4$.
The $N$ dimensional version of this model (i.e $x=(x_1,x_2,,x_N)$ ) was
solved in the limit $N \to \infty$ in Ref. \cite{Cule}.

In this Section we study a space discretized version of Sinai's model
which can be easily studied numerically. It is expected that this
discretized version has the same large time physics as the
continuous model (this was shown for some quantities
in the weak disorder limit \cite{result_1d_epl,bouchaud_1d}.
It is defined by a Fokker-Planck (FP)
equation:
 
 \beq \label{hopping}  \label{fplanck}
 \frac{d P_n(t)}{dt} = (H_{FP})_{nm} P_m(t) =
 W_{n,n+1} P_{n+1}(t) + 
 W_{n,n-1} P_{n-1}(t) - ( W_{n+1,n} + W_{n-1,n} ) P_{n}(t) 
 \label{fokkerplanckd}
 \eeq

where $P_n(t)$ is the probability that a particle is at site
$n$ at time $t$ (with some initial condition at $t=0$).
The hopping rates $W_{n+1,n},W_{n-1,n}$ are quenched random variables.
It is convenient to parametrize these rates as
in Ref. \cite{bouchaud_1d} in the form:

 \beq
 W_{n-1,n} = e^{- \phi_n}~~~ W_{n,n-1} = e^{\phi_n}
 \eeq
 
 This describes effectively the Arrhenius diffusion of a particle
 on a one dimensional lattice in a random potential
 $U_n = - 2 \sum_{k=0}^n \phi_k$. There is in effect a 
 force $2 \phi_n$ on the link between site $n-1$ and $n$.
 The equilibrium solution
 of equation \ref{hopping} corresponding to
 zero link current $J_{n,n-1}= W_{n,n-1} P_{n-1} - W_{n-1,n} P_{n} =0$
 is $P_n^{eq} = e^{-U_n}$. Here temperature is set to $T=1$.
 In the discrete Sinai model the variables $\phi_n$ are chosen
 gaussian independent from site to site,
 with $\overline {\phi_n \phi_{n'}}=\sigma/4 \delta_{nn'}$
and the random potential thus follows a gaussian discrete random
 walk as a function of $n$, with correlations 
 $\overline{(U_n - U_{n'})^2} = \sigma |n-n'|$.

The important quantity to determine is the Green function 
 $P(n,t|n_0,t_0)$ ($t \geq t_0$) of the FP operator $H_{FP}$
in a given environment. It is defined as the solution of 
 (\ref{hopping}) with initial condition $P(n,t_0|n_0,t_0)= \delta_{n n_0}$.
 
 We have computed numerically $P(n,t|n_0,t_0)$
 using exact diagonalization of the corresponding Schr\"{o}dinger operator
on a finite size ring. This 
 operator is a tridiagonal symmetric
 matrix which can be easily diagonalized for large size $L$.
The method, as well as the expressions for the correlation
 and response functions, is detailed in the
 Appendix \ref{appendixa}. We use $L+1$ sites $k=0,L$ 
 with both reflexive and periodic boundary conditions. We have used up to $L=250$ sites
and averages over $10^3$ and $5~10^3$ disorder configurations.
Because Sinai's diffusion is so slow,
we were able to study times up to $10^{15}$ without spurious
effects (edges, precision). We used $\sigma=2$ in all simulations
and checked the consistency of the results with
several random number generators.

 From this exact expression one can average either one 
or a product of two of these Green functions to obtain 
 single time or two times quantities, respectively.
For finite times, translational averages (with e.g 
uniform initial measure in an infinite single environment)
and averages over disorder should coincide. Note that
this will not be correct 
in a very special
case studied below of a periodic environment.
We now describe the results and their interpretation using
simple arguments.
 
 \subsection{single time quantities}

Averaged single time quantities can be obtained from
the disorder averaged Green function $\overline{P(n,t|n_0,0)}$.
There are some exactly known results for this single time quantities
and we will thus start by comparing with these results (as a check of
our simulations).

In \cite{result_1d_epl} the averaged probability density
at the origin $\overline{P(n_0,t|n_0,0)}$ was computed exactly
in the continuum limit for all times $t$. It describes the
weak disorder universal behaviour (see \cite{bouchaud_1d}).
At large $t$ the complete scaling form for the
{\it averaged diffusion front}, i.e the distribution of the
variable $n/(\log t)^2$ was obtained in \cite{kesten} as
$\overline{P(n,t|0,0)} \simeq (\log t)^{-2} p[n (\log t)^{-2}]$.
The function $p(x)$ is (up to a constant rescaling):

\begin{eqnarray}
p(x) = \frac{2}{\pi} \sum_{k=0}^{+\infty} 
\frac{(-1)^k}{2 k +1} exp(- \frac{(2 k+1)^2 \pi^2}{8} |x| )
\end{eqnarray}

\begin{figure}
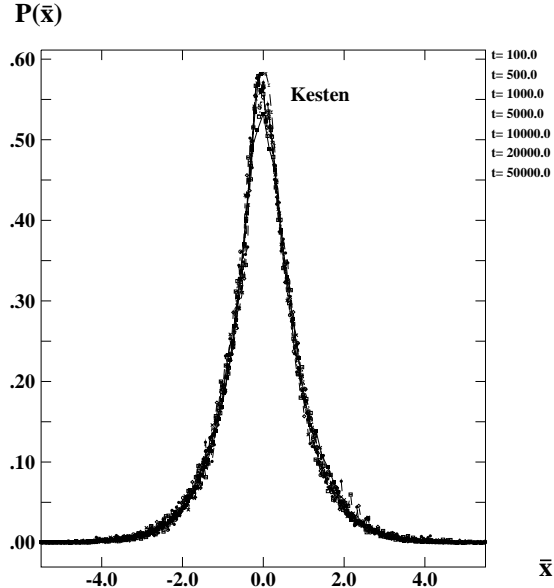


\label{fig1}

\centerline{\fig{8cm}{Kesten.eps}}

\caption{Single time averaged diffusion front: at large time
our simulations cannot be distinguished from
Kesten's prediction. We denote $\overline{x}=x/x_{rms}(t)$.
$L=125$ and $5 ~ 10^3$ configurations.}

\end{figure}

\begin{figure}
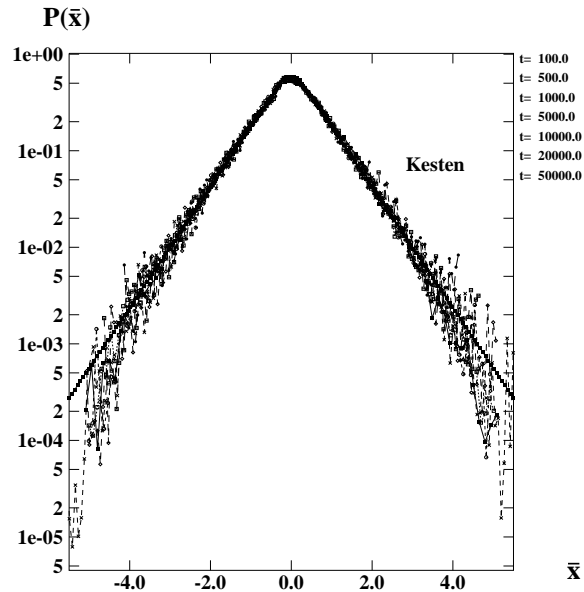


\label{fig2}

\centerline{\fig{8cm}{KestenLog.eps}}

\caption{Same as Fig. 1. in logarithmic scale for $P$}

\end{figure}

 We have computed numerically $\overline{P(n,t|0,0)}$ and the
 $x \sim (\log t)^2$ law as a check of our program. In Fig. 1
and Fig. 2
 we have plotted the scaling function $p[x]$ as it is determined 
numerically and the analytic result of Kesten (we have fixed the scale 
by imposing equality of the second moment and we denote
$x_{rms} = \sqrt{\overline{<x^2(t)>}}$ ). One can see that
the agreement is excellent. It improves
 considerably on an earlier determination by Nauenberg
\cite{nauenberg}. 
 The behaviour $x \sim (\log (t/t_0))^2$ is plotted in Fig. 3
 and also agrees. We have not attempted to fit the
amplitude to known results (we have checked that the order of magnitude 
is correct) since we were satisfied with the agreement in Fig. 1.

\begin{figure}
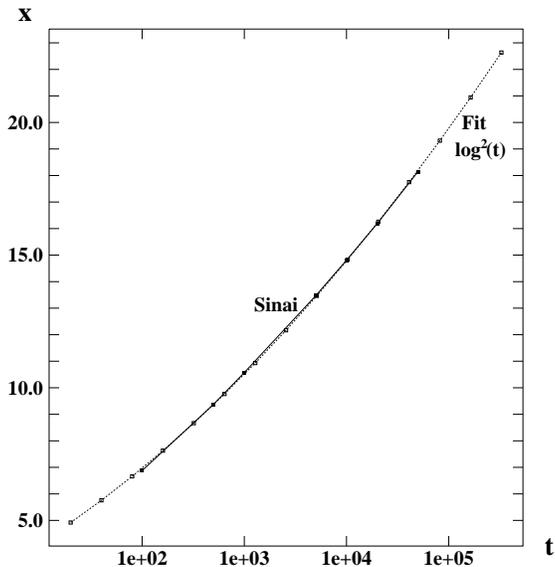


\label{fig3}

\centerline{\fig{8cm}{Kestenlaw.eps}}

\caption{Numerical data compared to Sinai's diffusion law $x \sim 
x_{rms}(t) \sim log(t/t_0)^2$.
$L=125$ and $5 ~ 10^3$ configurations.}

\end{figure}

We now turn to two time quantities.

\subsection{two time quantities}

There are no analytical results available at present, to our knowledge
for two time quantities. This make the numerical simulation all
the more important.

We have computed numerically the two time averaged Green function:
$\overline{ P(n,t|n',t') P(n',t'|n_0,t_0=0)}$. It is a complicated 
object which is difficult to analyze. Thus we have started by 
computing, as in mean field models, the configurationally
averaged mean squared displacement.
 
 \beqa  \label{btt}
 B(t,t') &=& \overline{ \langle (x(t)-x(t'))^2  \rangle } 
 = \sum_{n,n'} (n-n')^2 \overline{ P(n,t|n',t') P(n',t'|n_0,t_0=0)} \nn
 \eeqa

The best way is to plot it as a function of $t-t'$ for different 
waiting times $t_w=t'$. It is represented in Fig. 4
One sees that it clearly
does not depend only on $t-t'$ and that, the larger the waiting time,
the slower it grows. The dynamics slows down considerably
as $t' \to \infty$. This is very reminiscent of what happens in
mean field models. One also sees that an effective plateau develops
as $t'$ becomes larger. In mean field it would be a true
plateau \cite{Cule} which defines the Edwards Anderson order parameter $b_{EA}$.
Thus here an important question is whether there is a finite limit
for $B(t_w + \tau ,t_w)$ as $ t_w \gg \tau \gg 1$ as in
the mean field model.

\begin{figure}
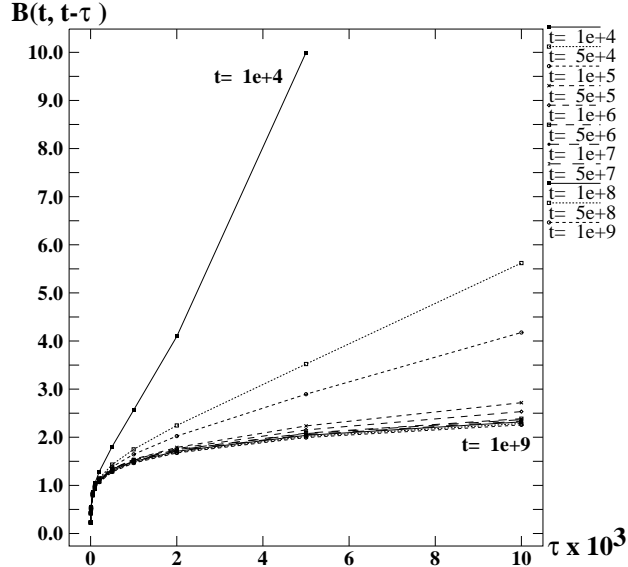


\label{fig4}

\centerline{\fig{8cm}{Bttw.eps}}

\caption{mean squared separation $B(t=t_w+\tau,t_w)$ as a function of $\tau$
for several $t \sim t_w$. A limit curve $B(\tau)$ can be seen, but $B(\tau)$ 
keeps growing with $\tau$ while in mean field it would go to a constant 
$b_{EA}=B(\tau=\infty)$. $L=125$ and $10^3$ configurations.}

\end{figure}

The answer to this question is {\it negative}. This is because, as will
be discussed below, in
finite dimension contrary to mean field, while the distribution
of $x(t)-x(t')$
converges towards a limit, it second cumulant is infinite.
In fact studying only the second cumulant $B(t_w + \tau ,t_w)$
is of little help in the present problem. One needs to study the
full disorder averaged distribution of relative displacements:

\begin{eqnarray}
\overline{Q(z,t,t_w)} = \int dx' \overline{ P(x'+z, t|x',t_w) P(x',t_w|0,t_0=0)}
\end{eqnarray}

and not only its second cumulant. We have found that when 
looking at that distribution, several regimes can be identified.

\subsection{quasi-equilibrium regime}

The first question to ask is whether there is a 
quasi equilibrium regime when $t$ and $t_w$ are close together.
It can be defined, for instance as
$t=t_w + \tau$ with $\tau$ fixed and finite (and 
$t_w \to \infty$). In mean field this would correspond to
the FDT regime, where time translational invariance
and the fluctuation dissipation theorem are found to hold
\cite{}.

One can thus ask whether the following limit distribution exists:

\begin{eqnarray}
\overline{Q(z,\tau)} = \lim_{t_w \to \infty} \lim_{L \to \infty} \overline{Q(z,t_w + \tau,t_w)}
\end{eqnarray}

where $L$ is the size of the system
and whether in that limit, some equilibrium theorems hold.

We are not able to answer to this important question
rigorously in all generality and we encourage other workers to do so. We will
however provide some elements which we hope
will shed some light on the issues.

One strategy is to first study the case of an {\it infinite periodic medium}
but with a {\it very large period} $L_0$ (i.e almost disordered). There
some things can be shown when $t_w$ is very large.
Of course this is cheating a bit since the idea would be to take
$L_0 \to \infty$ eventually, and thus this is like
interverting the limit $t_w \to \infty$ and $L_0 \to \infty$.
It does give some insight though. Thus we will then check, using numerics
and physical arguments, whether it can be extended to a non periodic case.

\subsubsection{a periodic model}

We are considering an infinite periodic medium. 
We assume that the potential $U(x)$ is periodic
which corresponds to the situation with no drift. One can generalize
to the case with a bias by taking a periodic force (using
the results of \cite{ledou_vino}.
We will first consider a single environment
(no configurational averages).

We are interested in the distribution:

\begin{eqnarray}
Q_{L_0}(z,t_w + \tau,t_w) = \int_{-\infty}^{+\infty} dx'
P_{L_0}(x'+z, t_w + \tau| x' t_w| x_0 0)
\end{eqnarray}

It can be written as a sum over periods:

\begin{eqnarray}
&& Q_{L_0}(z,t_w + \tau,t_w) = \sum_k \int_{0}^{L_0} dx'
P_{L_0}(x'+k+z, t_w + \tau| x'+k, t_w) P_{L_0}(x'+k,t_w|x_0 0) \\
&& = \int_{0}^{L_0} dx' P_{L_0}(x'+ z, \tau| x',0) \tilde{P}_{L_0}(x',t_w|x_0 0)
\end{eqnarray}

where we have defined the periodized Green function:

\begin{eqnarray}
\tilde{P}_{L_0}(x', t' | x_0 0) = \sum_{k} P_{L_0}(x'+ k L, t'| x_0 0)
\end{eqnarray}

and we have used the periodicity of
the disorder $P_{L_0}(x, t | x', t') = P_{L_0}(x+ m L, t | x'+m L,  t')$.
So, up to now it is exact for any configuration of the disorder.

Since the periodized function $\tilde{P}$ is in fact the Green function
of the problem on a ring of size $L_0$ (see e.g \cite{derrida,ledou_vino}),
it converges towards the equilibrium Gibbs measure on the ring
in the large $t_w$ limit:

\begin{eqnarray}
\lim_{t_w \to \infty} \tilde{P}_{L_0}(x', t_w | x_0 0) =
\frac{e^{- U(x')/T}}{\int_{0<x'<L_0} dx' e^{- U(x')/T}} 
\end{eqnarray}

This implies that:

\begin{eqnarray}  \label{period1}
Q_{L_0}(z,\tau) \to \int_{0}^{L_0} dx' 
P_{L_0}(x' + z, \tau |x' 0) \frac{e^{- U(x')/T}}{\int dy e^{- U(y)/T}}
\end{eqnarray}

This is particularly useful when $\tau$ is fixed and much smaller
than the time necessary to travel a period (i.e $(\ln \tau)^2 \ll L_0$ in
Sinai's model). This formula can also be generalized in presence
of a drift using the stationary distribution obtained in
\cite{ledou_vino}.
Note that these arguments extend to any finite dimensional problem.

\subsubsection{non periodic case}

In the non periodic case, when $L_0 \to \infty$ before $t_w$,
it is obvious that for a single configuration of disorder there
is no limit to $Q(z,t_w +\tau, t_w)$ as $t_w \to \infty$. Indeed
numerically Sinai's diffusion consists of sudden jumps to deeper and
deeper wells. Thus even though in each successive well there is presumably a
quasi-equilibrium regime $Q(z,t_w +\tau, t_w) \sim Q_{\text{well}}(z,\tau)$
it will depend on the details of each
new well encountered. In some sense there is a {\it distribution} of such 
quasi equilibrium distributions (which is not unlike the image from replica 
symmetry breaking). In addition, there are times $t_w$ where the packet jumps,
but they presumably become more and more rare at large $t_w$.
Thus only the disorder averaged (or translational
average) can be expected to converge to some limit $\overline{Q(z,\tau)}$,
as these features will get smoothed out.
We will assume that this convergence holds (based on our numerical evidence).

One possible further assumption is that once averaged over disorder one has:

\begin{eqnarray} \label{convergent}
\overline{Q(z,\tau)} = \overline{Q_{eq}(z,\tau)} \equiv \lim_{L_0 \to \infty} \overline{Q_{L_0}(z,\tau)}
\end{eqnarray}

where $Q_{L_0}(z,\tau)$ is the distribution in the periodized medium
(using the same disorder distribution) given in (\ref{period1}). We have called
$Q_{eq}(z,\tau)$ this distribution.

A possible reasonable starting assumption is that they are equal in 
$\overline{Q} =\overline{Q_{eq}}$ in Sinai's model. We do have good numerical indication
that this is indeed correct, as shown in Fig. 5. However, because
differences, if they exist, could be subtle, small and hard to detect
we emphasize that whether this is strictly correct should be
checked further (there are indeed cases where it is wrong - see 
the solvable model of Section \ref{solvable}).
In any case, the above equation provides at least a basis for comparison.

\begin{figure}
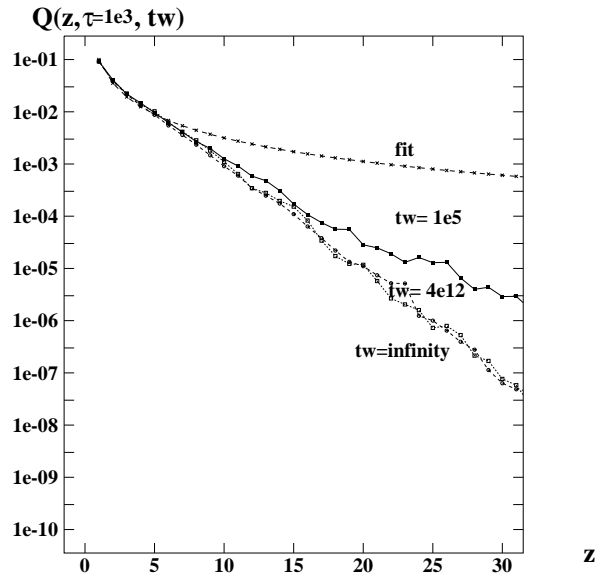


\label{fig5}

\centerline{\fig{8cm}{Qztautw.eps}}

\caption{Convergence, for a fixed $\tau$, of $\overline{Q(z,t_w+\tau,t_w)}$ (represented here
for $t_w=10^5$ and $t_w=4~10^{12}$) towards $\overline{Q_{eq}(z,\tau)}$
computed here from formula (\ref{period1},\ref{convergent})
 (curve called ``$t_w=$ infinity'' in the figure).
The limit curve $\overline{Q(z,\tau=\infty)}$ is shown by comparison (curve called ``fit'').
 $L=200$ and $5~10^3$ configurations.}

\end{figure}

\subsubsection{correspondence with the statics}

Let us investigate the large $\tau$ limit of $\overline{Q(z,\tau)}$.
One expect, if the above is assumption is correct, that:

\begin{eqnarray} \label{period}
\overline{Q(z,\tau)} \sim \lim_{L_0 \to \infty} \overline{Q_{L_0}(z,\tau)}
\to \lim_{L_0 \to \infty} \overline{ Q_{L_0}(z) } = 
\lim_{L_0 \to \infty} \overline{ \int_{0}^{L_0} dx' \frac{e^{- U(x'+z)/T} e^{- U(x')/T}}{(\int dy e^{- U(y)/T})^2} }
\end{eqnarray}

Thus the two time calculation becomes a two replica correlation
function in a replica calculation.

\medskip

{\it properties of the quasi static distribution}

\medskip

For Sinai model one can argue that there is indeed a limit 
averaged quasi static distribution $\overline{Q}(z)$. 
From the above formula (\ref{period}) and since
there is typically one global minimum which dominates the period
$L_0$ in Sinai's landscape, it is clear that this distribution
will consist of two parts:

(i) a part localized near $z=0$ of finite extent, with a
finite weight.

(ii) a part which is strongly non self averaging
 with, for each environment, some localized peaks around
some $z_i$. Averaging over disorder yields a smooth
$Q(z)$ with {\it algebraic tail} at large $z$:

\begin{eqnarray} \label{tail}
Q(z) \sim \frac{A}{z^{3/2}} 
\end{eqnarray}

This tail comes from configurations of the disorder which look
as in Fig. 6. The lowest well $U_{min}$ in the sample of period $L_0$ is
represented and it happens that there is a secondary well, with
a bottom at $U_{min} + E_n$ with $E_n = O(T)$, at a finite distance 
$z_0$ of the first one. Then roughly the measure $Q(z)$ will consist of
two peaks localized around the two wells.

Since in Sinai model the random potential can itself be seen
as performing a random walk, the probability that such an environment
occurs can be estimated from the probability of return to the origin
of a random walk. This yields a probability $\sim z_0^{- 3/2}$
and thus the above algebraic tail of $Q(z)$. More refined calculation,
taking into account that the principal well is an absolute minimum can 
be performed and lead to estimates for $A$, but go beyond this paper.
Let us point out that the
above equilibrium Gibbs measure has been analyzed recently in Ref. \cite{brodericks}
and they also obtained analytically the above algebraic tail
(i.e formula (\ref{tail})).
This tail however can be explained from simple arguments.

It would be interesting to check whether one can
extend ideas from mean field and whether
in a replica calculation of (\ref{period})
 the localized part near the origin
would correspond to the replica symmetric part of the solution while the tails
from the rare events to the part with broken replica symmetry.

So there is also an Edwards Anderson order parameter here
but it is a {\it distribution}. Because of non self averaging effects
one needs the full distribution $Q(z)$. It would be nice to
check more rigorously that $\overline{Q(z)} = \overline{Q_{eq}(z)}$ (and thus
that there is no extra {\it dynamical} order parameter).

Finally let us emphasize that the above arguments for the tail
of the quasi equilibrium distribution are very similar to static droplet model
arguments \cite{huse_fisher}. Here however they are made in a dynamical
context.

\begin{figure}

\label{fig6}

\centerline{\fig{8cm}{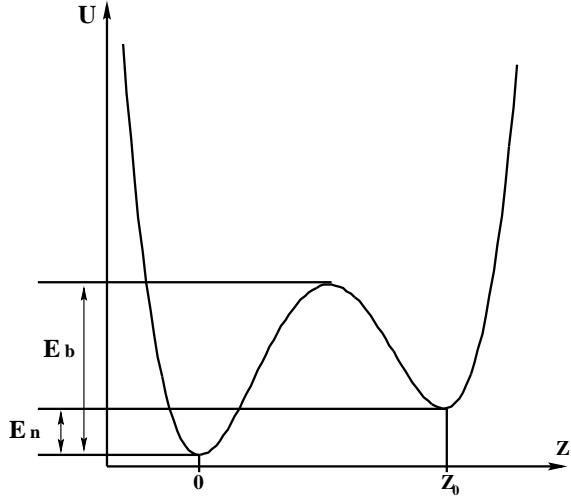}}

\caption{Two well (FDT-TTI) quasi equilibrium regime: rare configurations
leading to sample fluctuations and algebraic decay of $Q(z)$}

\end{figure}

The above arguments thus suggest that the moments of the relative displacements:

\begin{eqnarray}  \label{mom}
\lim_{t_w \to \infty} \overline{< (x(t_w + \tau) - x(t_w))^n >}
\sim \int z^n Q(z,\tau) 
\end{eqnarray}

converge towards a finite limit when $\tau \to \infty$ only
for $n< 1/2$. For $n>1/2$ they grow unboundedly with $\tau$.
These moments growing with $\tau$ may not a priori be incompatible with
being in a quasi equilibrium state (though it calls for 
further investigations). In mean field, this does not happen
since even for unbounded diffusion problems \cite{Cule} the quasi equilibrium regime is 
such that $B(t,t') < b_{EA}$.

We now estimate these moments (\ref{mom}) using a simple two well
model taking into account the crossing of the barrier $E_b$ between the
two wells. Let us write:

\begin{eqnarray} \label{simplemodel}
Q(z,\tau) = \delta(z) \frac{1}{2} (1 + e^{-\tau e^{-E_b}}) + 
\delta(z-z_0) \frac{1}{2} (1 - e^{-\tau e^{-E_b}})
\end{eqnarray}

and average over $z_0$ and the barrier height 
which we can take to scale as $E_b(z_0) \sim \alpha z_0^{1/2}$
$\alpha$ being a positive random variable. This yields 
the moments:

\begin{eqnarray}
\lim_{t_w \to \infty} \overline{< (x(t_w + \tau) - x(t_w))^n >} \sim
\frac{1}{2} \int dz z^n \frac{A}{z^{3/2}} (1 - e^{-\tau e^{-\alpha z^{1/2}}} )
\end{eqnarray}

Introducing the scaled variable $y = z/\ln^2 \tau$ and using that
the function $(1 - e^{- e^{- \alpha \ln^2 \tau (y^{1/2} -1/\alpha^2)}}) \approx
\theta(1/\alpha^2 -y)$ one finds, for $n>1/2$. 

\begin{eqnarray}  \label{moments}
\lim_{t_w \to \infty} \overline{< (x(t_w + \tau) - x(t_w))^n >}
\sim A (\ln \tau)^{2 n -1}
\end{eqnarray}

\begin{figure}
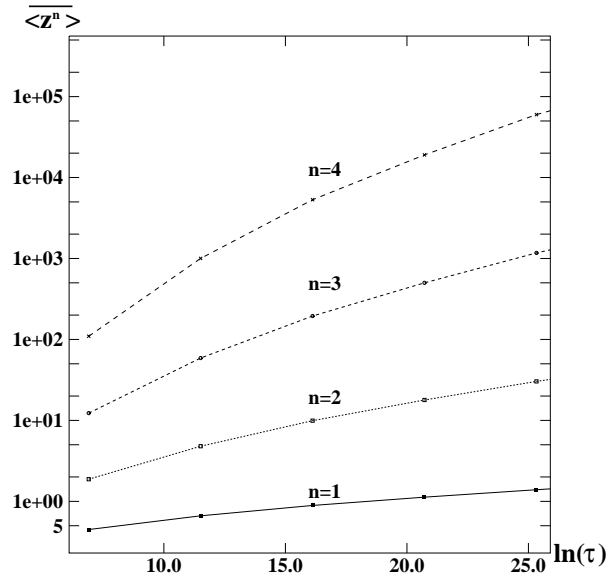


\label{fig7}

\centerline{\fig{8cm}{MomentsQ.eps}}

\caption{Numerical determination of the moments $<|z|^n(\tau)>_Q$ as a function
of $\ln \tau$. $L=200$ and $5~10^3$ configurations.}

\end{figure}

We have plotted these moments as determined numerically in
Fig. 7. The logarithmic growth of moments $n>1/2$ is
clearly demonstrated (we have checked that e.g. $n=1/4$ saturates at
large $\tau$). Though they seem to follow (\ref{moments})
qualitatively, a more quantitative agreement is probably difficult to reach
numerically - since these effects entirely come from rare events !

The above considerations also predict that the distribution $\overline{Q(z,\tau)}$ 
should converge towards $\overline{Q(z)}$ with a $z/\ln^2 \tau$ scaling behaviour.
This is consistent with our simulations as can be seen in
Fig. 8 where we have shown the $\overline{Q_{eq}(z,\tau)}$
for various $\tau$. It does indeed converge towards 
a limit curve $\overline{Q_{eq}(z)}$ which was well fitted by
its asymptotic behaviour $\sim z^{-3/2}$.

One can also check from this simple model (\ref{simplemodel}) that as a consequence,
the Edward Anderson ``overlap'' parameter $\overline{Q(z=0,\tau)}$
should also converge towards it equilibrium
value as $\tau \to \infty$ with $1/(\ln \tau)^a$ corrections.
This indeed happens as is shown in Fig. 9.

\begin{figure}
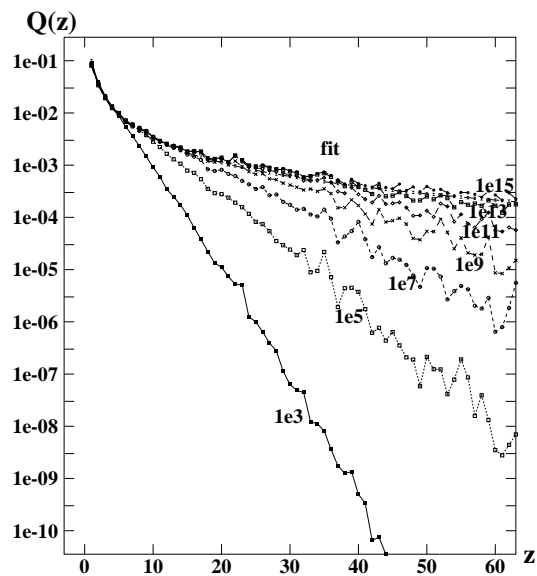


\label{fig8}

\centerline{\fig{8cm}{Qz.eps}}

\caption{Numerical determination of $\overline{Q_{eq}(z,\tau)}$ compared to the
predicted asymptotic behaviour $\sim 1/z^{(3/2)}$ (fit).$L=200$ and $5~10^3$ configurations.}

\end{figure}

\begin{figure}
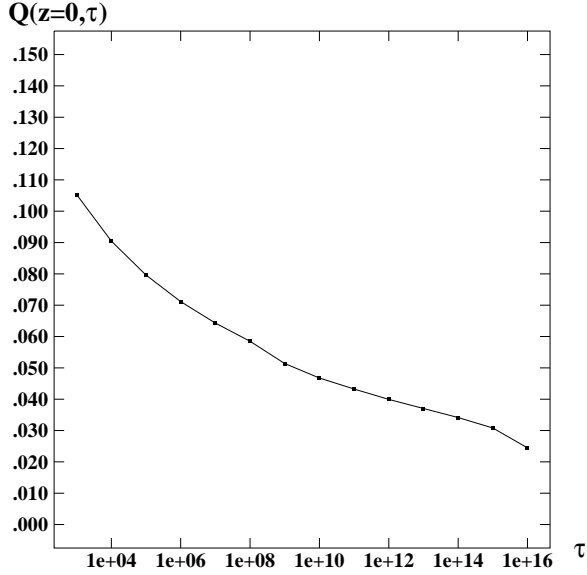


\label{fig9}

\centerline{\fig{8cm}{Qz0tau.eps}}

\caption{Plot of $\overline{Q_{eq}(z=0,\tau)}-\overline{Q_{eq}(z=0,\tau=\infty)}$
against $\tau$ (in $\ln \tau$ scale).
$L=200$ and $5~10^3$ configurations.}

\end{figure}

\subsubsection{ equilibrium theorems}

It is plausible that for large $t_w$ equilibrium theorems like FDT 
hold for disorder averaged quantities. These provide relations between
correlation functions and linear response functions.

It is useful to generalize these theorems to the full
probability distribution $\overline{Q}(z,t,t')$. This is done in the Appendix
\ref{appendixa} and \ref{appendixb}
to which we refer for details. One defines:

\beq
\overline{R(z,t,t')} = \frac{\delta \overline{Q_h (z,t,t')}}{\delta h(t')}
\eeq

where $\overline{Q_h (z,t,t')}$ is the probability when 
an additional infinitesimal uniform field pulse $h(t)=h \delta(t-t')$ is applied
at time $t'$. A technical detail is that it can be applied at time $t'-\epsilon$ 
(which defines $R^{+}(z,t,t')$) or at time $t'+\epsilon$
(which defines $R(z,t,t')$ and corresponds to the Ito prescription
for the response, i.e $\delta<x(t')>/\delta h(t')= 0$). 
If quasi equilibrium hold
(meaning the current $J$ at $t_w$ vanishes, see
Appendix \ref{appendixb}, the time translational invariant averaged functions  
$\overline{Q}(z,\tau)$ for large $t'=t_w$
and fixed $\tau$ and $\overline{R}(z,\tau)$ should verify the
exact differential relation:

\beq   \label{itoeq}
 - \partial_{\tau} \overline{Q(z,\tau)} =  T \partial_z \overline{R(z,\tau)}
 \eeq

These relations can be generalized to discrete models as
done in \ref{appendixa}.
The response $R^{+}(z,\tau)$ satisfies a slightly different equation:

\beq
 - \partial_{\tau} \overline{Q(z,\tau)} = - T \partial_z^2   \overline{Q(z,\tau)}
 + T \partial_z \overline{R^{+}(z,\tau)}
 \eeq

We determined both the distribution $\overline{R^{+}(z,\tau)}$ 
and $\overline{R(z,\tau)}$ numerically
(as explained in \ref{appendixb} ). The response $\overline{R^{+}(z,\tau)}$
is plotted in Fig. 10.
Note that for $\tau=\infty$ the above implies simply:

\beq
\overline{R^{+}(z,\tau)} = \partial_z \overline{Q(z,\tau)}
\eeq

and we checked that $\overline{R^{+}(z,\tau)}$ has the form $\sim 1/z^{(5/2)}$
expected from the above relation and (\ref{tail}).

\begin{figure}
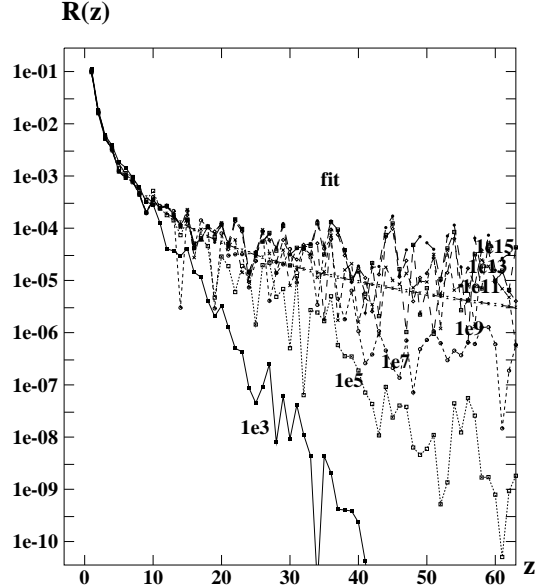


\label{fig10}

\centerline{\fig{8cm}{Qhz.eps}}

\caption{Plot of $\overline{R^{+}(z,\tau)}$ ($y$ coordinate)
versus $z$ for various $\tau$ and comparison with the expected
large $\tau$ limit $1/z^{(5/2)}$ (``fit''). $L=200$ and $5~10^3$ configurations.}

\end{figure}

The Ito response $ \overline{R(z,\tau)}$
satisfies (\ref{itoeq}) and is thus analogous
to a {\it probability current}. Examining $ \overline{R(z,\tau)}$ obtained from
our simulation as a function of
$z$ confirms the above arguments based on the two well model:
there is a (statistical) current flowing from
the center region $z\sim 0$ to the $z>0$ (and to the $z<0$).
In each disorder configuration $R(z=0,\tau)$ does not have to vanish
(since a each local environment is not symmetric)
but its average  $ \overline{R(z=0,\tau)}$ must vanish by symmetry.
We observe that in the simulation it is of the order $10^{-3}$ smaller
than $\overline{R(z=1,\tau)}$ which is consistent.

We have plotted $ \overline{R(z=5,\tau)}$ in Fig. 11. It shows clearly 
a $1/\tau$ decay. To be more accurate, we have plotted
$\tau \ln \tau \overline{R(z=5,\tau)}$
in Fig. 12. It indicates that:

\beqa
\overline{R(z,\tau)} \sim \frac{1}{\tau (\ln \tau)^b}
\eeqa

with $b \approx 1$ (for a fixed $z$). This could be consistent with a decay
of $Q(z,\tau) \sim 1/\ln^2\tau$ for fixed $z$ and 
a scaling $z \sim \ln^2 \tau$ in (\ref{itoeq}). However,
$Q$ and $R$ are expected to consist of two
parts with $\tau$ dependent relative weights, one
part scaling as $z/\ln^2 \tau$ and a fixed part in
$z$. Thus more work is needed to determine these functions
more precisely.

\begin{figure}

\label{fig11}

\centerline{\fig{8cm}{Rztau.eps}}

\caption{Plot of $\overline{R(z=5,\tau)}$ ($y$ coordinate)
as a function of $\tau$. Two simulations are indicated:
$L=100$ and $5~10^2$ configurations and
$L=200$ and $10^3$ configurations}

\end{figure}

\begin{figure}
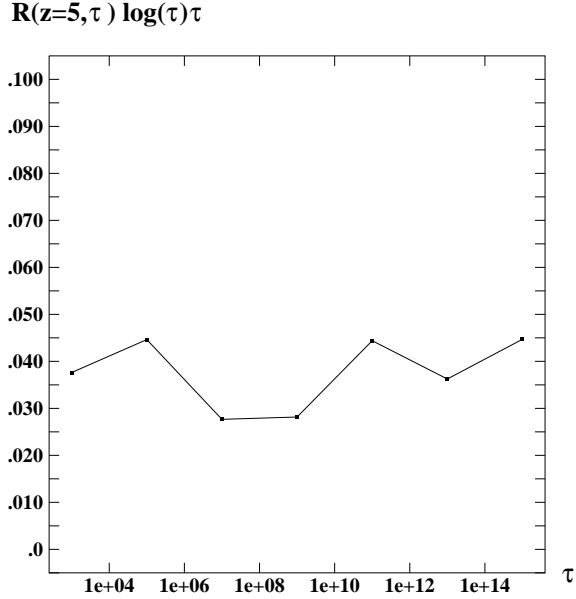


\label{fig12}

\centerline{\fig{8cm}{TauLogRz.eps}}

\caption{Plot of $\tau \log(\tau) \overline{R(z=5,\tau)}$ ($y$ coordinate)
as a function of $\tau$. $L=200$ and $10^3$ configurations.}

\end{figure}

Very recently (while this manuscript was in completion) a
promising approach was developped by Cugliandolo, Dean and Kurchan
(CDK) \cite{Cudeku} to obtain bounds which may permit to demonstrate that
these quasi-equilibrium regimes exist in models such as Sinai's.
Their approach is explained in Appendix \ref{appendixc} and some extensions
and applications are given. For Sinai's model the bounds should
be performed on disorder averaged quantities (since single environment
ones do not converge). Doing this one can obtain two bounds of
interest for Sinai's model:

\beqa  \label{bound1}
| \overline{<x^2(t)>} - \overline{<x(t) x(t')>}
- T \int_{t'}^{t} R(t,t')| \leq (\overline{<x^2(t)>})^{1/2} \int_{t'}^{t} ds
( |\frac{d \overline{H}(s)}{d s}|^{1/2} 
\eeqa

and:

\beqa \label{bound2}
| \partial_{t'} B(t,t') + 2 T R(t,t') | < B(t,t')^{1/2} 
|\frac{d \overline{H}(t')}{d t'}|^{1/2}
\eeqa

The first one was given in \cite{Cudeku}
(though the problem of disorder averaging is not
discussed) while the second is new. $B(t,t')$ is defined in
(\ref{btt}) and $R(t,t')=\delta \overline{<x(t)>}/\delta h(t')= \int dz z \overline{R(z,t,t')}$
(with Ito's definition of the response). 

The function $\overline{H}(t') = \overline{ 
\int dx' P(x't'|0 0) (T \ln P(x',t'|0 0) - U(x')) }$ is the averaged
free energy which satisfy an $H$ theorem (it is always decreasing).
The nice observation of \cite{Cudeku}
is that since in Sinai's model $\overline{H}(t')  \sim - \ln t'$
the above bound (\ref{bound1}) implies that the r.h.s bound for the integrated FDT violation
is $ (\ln t)^2 (t^{1/2} - {t'}^{1/2})$. Clearly for $\tau=t-t'$ fixed it implies
that (integrated) FDT holds, but it does even seem to imply that it holds further 
even for $\tau \sim t'^{1/2}$.

Note that the second bound (\ref{bound2}) immediately implies
that if there is a limit distribution
$\lim_{t_w \to \infty} \overline{Q (z,t_w+\tau,t_w)}$ then it will depend
only on $\tau=t-t'$ (TTI) and the bound imply that FDT is verified
(since $dH(t')/dt'$ goes to zero). (Note that this is also
the case for the Brownian motion).

While these results are as inescapable as rigorous bounds, 
an explanation for this could be looked for in our previous arguments
about periodic media. Indeed if our assumption (\ref{convergent}) is correct,
then one can choose $\ln^2 \tau \ll L_0 \ll \ln^2 t_w$ and still
have equilibrium and FDT (since the probability of having another
accessible absolute minimum in the box $L_0$ 
at distances $\gg 1$ is vanishingly small (see )). Thus 
it is likely that in fact FDT will hold beyond what is 
shown by the bound, probably until
$\tau \sim t_w^c$ with $c <1$ (beyond that one enters
the diffusion and aging regime, see next section).

Note also for completeness the two single time bounds:

\beqa
&& |\partial_{t'} \overline{<x(t')>}| \leq |\frac{d \overline{H}(t')}{d t'}|^{1/2} \\
&& |\partial_{t'} \overline{<x(t')^2>}| \leq |\overline{<x(t')^2>}|^{1/2} |\frac{d \overline{H}(t')}{d t'}|^{1/2}
\eeqa

Thus we have found that an (equilibrium) TTI diffusion regime for the
process $z(\tau)$ persists within the quasi equilibrium regime. 

One can also apply these bounds to the case with a an applied
force $f \sim \mu$, anticipating a little on the next Section \ref{directed}
where the directed model will be discussed.
Let us assume that $\overline{H}(t') \sim - <x(t)> \sim - (t')^\mu$.
Then (\ref{bound1}) leads to:

\beqa  \label{bound10}
| \overline{<x^2(t)>} - \overline{<x(t) x(t')>}
- T \int_{t'}^{t} R(t,t')| \leq t^\mu (t^{(\mu + 1)/2} - (t')^{(\mu + 1)/2})
\eeqa

Expanding for $\tau \ll t'$, one finds that the l.h.s must be smaller than 
$\tau/t'^{(1-3 \mu)/2}$. Thus it shows the existence of an FDT regime
for $\mu >0$ in Sinai's model. The r.h.s goes strictly to zero
as long as $\mu < 1/3$ but if one actually divide the r.h.s by the
expected scale of the l.h.s ($\sim t^{2 \mu}$) one gets
$\tau/t'^{(1+\mu)/2}$. Thus a FDT regime (with $X=1$) seems to be possible
for $\mu <1$.

This may seem surprising at first, since for $\mu>0$ the system is
driven. However it makes sense physically. Remember that
for $\mu <1$ the particles will spend most (all for $t_w \to \infty$)
of their time in
a well of release time $\sim t_w$ (see section \ref{directed}).
There they have time to equilibrate.
Thus in some sense the fact that there is aging (and broad distributions
of release time) is intimately related to the fact that there
can be equilibrium within a well.

The above bounds also put constraints on the possible aging forms
for a large class of models. This will be discussed in Section
(\ref{barriers}).

\subsection{Aging and diffusion regime}

We will now present our results for 
$\overline{ P(n,t|n',t') P(n',t'|n_0,t_0=0)}$ in the regime 
where $t$ and $t'$ are large and well separated. As stressed above
the data is complicated to analyze. We have found numerical
evidence for two regimes.

\medskip

(i) {\it the diffusion regime}

\medskip

The first one is the {\it diffusion
regime} where $z \sim (\ln t)^2$ and $\ln t' \sim \ln t$:

\begin{eqnarray}  \label{twotime}
\overline{Q(z,t,t')} \sim \frac{1}{(\ln t)^2} F[ \frac{z}{(\ln t)^2} , \frac{\ln t'}{\ln t} ]
\end{eqnarray}

In that regime the three relative displacements
$x(t)-x(t')$, $x(t)-x(0)$ and $x(t')-x(0)$ are of the same
order of magnitude.

\medskip

(i) {\it the aging regime}

\medskip

Another regime was found by looking at the decay of $Q(z \sim 0,t,t')$
for $z$ fixed and small (near $z=0$) and $t$ and $t'$ large.
By $z$ fixed and small we mean in our numerical simulation $n=n'$ but
it means more generally that $n$ is in a finite neighborhood of $n'$.
For instance we have found that
the decay of $Q(z=0,t,t')$ could be as slow as desired by taking
both $t$ and $t'$ to infinity. We have tried various dependences
of the form:

\begin{eqnarray}
\overline{Q(z=0,t,t_w)} \sim f[\frac{h(t_w)}{h(t)} ]
\end{eqnarray}

The best fit was obtained for:

\begin{eqnarray}
h(t) \sim \ln t
\end{eqnarray}

This is illustrated in Fig. 13 where we have used $t_w \sim t^{a}$
with $0<a<1$ and varied $t$ with fixed $a$ which is consistent with
$h(t) = \ln t$, since $a=\ln t_w/\ln t$. For $a=1$ one recovers
$\overline{Q_{eq}(z=0)}$ as expected. 
Note that it is more difficult numerically to get correctly the
small $a$ behaviour (hence the curvature of the curves
for small $a$ on Fig. 13 which is due to the 
smallness of the available $t_w \sim t^a$). 
We have also tried $t_w \sim t/2$ in order to test a possible $t/t_w$ dependence
of this quantity (upper curve). This curve is in fact indistinguishable
at large $t$ from $\overline{Q_{eq}(z=0)}$. This strongly indicate that
aging with a form $t/t_w$ is unlikely.

We want to emphasize that we could not rule out other forms
more complicated than $h(t)=\ln t $. However, the form $t/t_w$ appears very unlikely
for the measured quantity (e.g we haven't checked the behaviours
of the moments).

\begin{figure}
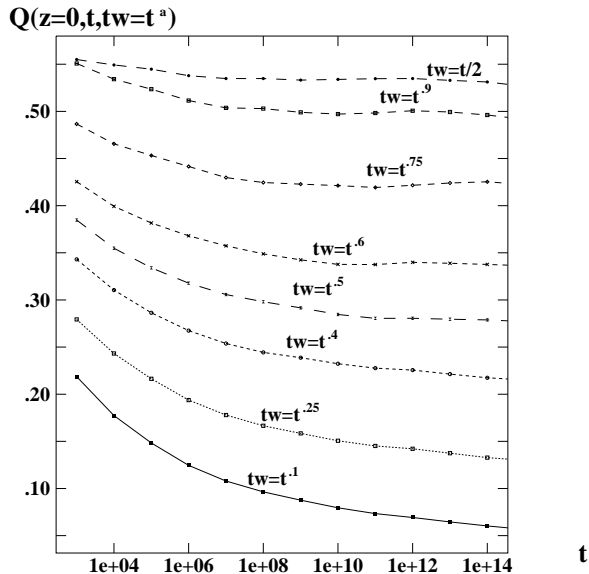


\label{fig13}

\centerline{\fig{8cm}{Qz0ttw.eps}}

\caption{Plot of $\overline{Q(z=0,t,t^a)}$ for various values of $a$ which
indicates that the aging scaling function is $h(t) \sim \ln t$.
The curvature of the curves for small $a$ at times not very large
is due to the smallness of $t^a$. $L=200$ and $5~10^3$ configurations.}.

\end{figure}

We have also computed $\overline{Q(z,t,t^a)}$ for several $z$ fixed and large $t$.
We find that the aging behaviour is also valid for $z \neq 0$.
In fact we find strong evidence for the behaviour:

\beqa 
\overline{Q(z,t,t')} = Q_0(z) f[\frac{\ln t'}{\ln t}]
\eeqa

as is shown in Fig. 14 for $0 \leq z \leq 10$, where we have been
able to collapse all the various curves $\overline{Q(z,t,t')}$ on
the same curve $Q_0(z)$. Thus this curve should also be equal to
$Q_0(z)= Q_{eq}(z,\tau=\infty)/Q_{eq}(z=0,\tau=\infty)$.

This behaviour seems to be consistent with a picture
of aging in Sinai's model (at least for small $z$) 
where an equilibrated packet jumps as a whole out of its
large well when $t' \sim t^a$. This will be discussed further in
Section \ref{barriers}. Though the collapse in Fig. 14 is perfect at
small $z$ (better than $10^{-4}$ in relative precision) it becomes
poorer at larger $z$. This is probably due to the fact that for the available times
one overlaps somehow with the diffusion regime. It is unlikely,
but not ruled out, that this could be the sign of yet another regime
(more work would be necessary).

\begin{figure}

\label{fig14}

\centerline{\fig{8cm}{QzFcscal.eps}}

\caption{Plot of $\overline{Q(z,t,t^a)}/\overline{Q(z=0,t,t^a)}$ as a function
of $z$ for several values of $a$ and for two times $t=10^{10}$ and $t=10^{13}$.
$L=250$ and $10^3$ configurations.}

\end{figure}

The fact that the temporal scaling is found to be 
the same $h(t) \sim L(t) \sim \ln t$ in both this
aging regime and the diffusion regime indicate
that these two regimes may in fact be the same in this
model. If that is indeed correct it does imply that the
above two time diffusion scaling function (\ref{twotime}) must have a
{\it singularity near the origin}, e.g a delta function singularity.
This remains to be investigated further.

Note that in mean field, in short range models, one can also see
that the diffusion scale and the aging scales actually
coincide \cite{Cule}. Here it seems that the same happens, though for a different
quantity $Q(z \sim 0,t,t')$.

 \section{Directed Model} \label{directed}

In this Section we study aging and diffusion in a simple model of directed diffusion
with disorder where two time correlation functions 
as well as response functions can be computed simply.
This model will appear as a particular case of a more general
`solvable model' which we will introduce in Section \ref{solvable}

In this model the particle can only jump to the right of the
 occupied site and quenched disorder is introduced by choosing at
 each site $n$ an average waiting time $\tau_n$ according to a 
 given distribution $P(\tau)$. The distribution of waiting times 
 is further chosen with an algebraic tail at large times 
 $P(\tau) \sim C/\tau^{1+\mu}$.
This directed model was introduced and studied (for single time
quantities) in \cite{pld_thesis}. It was also 
shown \cite{pld_thesis,bouchaud_1d} that it is a large scale {\it effective description}
of Sinai model in presence of a bias. The idea is that
in presence of an applied force $f$ the potential landscape $(x, U(x))$ 
is a biased random walk. There are thus some places where the walker goes back against
the bias. This leads to rare barriers against the drift of size $E_b$ with
probability $\exp(- f E_b)$. Since the waiting time in these traps
behaves as $\exp(-E_b/T)$ this is enough to generate dynamically
an algebraic distribution of waiting times \cite{bouchaud_1d}.
Whether this model is also a good description of the biased Sinai model
for two times quantities remain to be investigated in details. 

The model is defined by the Fokker-Planck (FP) equation:
 \beq
 \frac{d P_n(t)}{dt} = H_{nm} P_m(t) \equiv
 W_{n-1} P_{n-1}(t) - W_{n} P_{n}(t) 
 \label{fokkerplanck}
 \eeq
 The rates at each site correspond to a mean waiting
 time $\tau_n = 1/W_n$.

The Green function $P(n,n_0|t-t_0)$ is defined as the solution of 
 (\ref{fokkerplanck}) with initial 
 condition $P(n,n_0|0)= \delta_{n n_0}$. Its Laplace Transform (LT) 
$P(n,n_0|s) = \int_0^{+\infty} P(n,n_0|t) e^{- s t}$ can be computed
easily from (\ref{fokkerplanck}). It reads:

 \beq
 P(n,n_0|s) = \frac{1}{s+W_n} \prod_{k=n_0}^{n-1} 
 \frac{W_k}{s+W_k}
 \eeq
 
Averaged single time averaged quantities are easily computed
from:

 \beqa
 && \overline{P(0,0|s)} = \overline{\frac{1}{s+W}} \equiv \Phi(s) \\
&& \overline{P(n,0|s)} =  \Phi(s) (1 - s \Phi(s))^{n-1}
 \eeqa

The long time limit is then obtained from the small $s$ behaviour
of the function $\Phi(s)$:

 \beq
 \Phi(s) \sim C \pi / (\sin \mu \pi) s^{\mu-1}  ~~~~ \mu < 1
 \eeq 
 \beq
 \Phi(s) \sim 1/V + C \pi / (\sin \mu \pi) s^{\mu-1}  ~~~~~ 1< \mu < 2
 \eeq 
 \beq
 \Phi(s) \sim 1/V - D s + C \pi / (\sin \mu \pi) s^{\mu-1} ~~~~~ \mu > 2
 \eeq 

This yields the several phases of the model: the subdiffusive phase
$0< \mu < 1$ where $ x \sim t^\mu$ (and zero velocity), the anomalous dispersion phase
$1< \mu < 2$ where there is a velocity $V >0$ but dispersion is 
anomalous $D=\infty$, and the diffusive phase $\mu > 2$. We will be mostly
interested here in the case $0< \mu <1$. There the above result immediately
yields a Levy diffusion front \cite{pld_thesis}.

We are now interested in the averaged probability that the particle
advances by $m$ between $t'$ and $t=t'+\tau$:

 \beq
 Q(m,\tau,t') = \langle \overline{ \delta( x(t) - x(t') - m )}
 \rangle = 
 \sum_{n \ge 0} \overline{ P(n+m,n,\tau) P(n,0,t') }
 \eeq
 
 This quantity is computed in the Appendix \ref{appendixf}. The result in Laplace 
variables is the following:

 \beqa
 Q(m=0,s_1,s_2) &=&  \frac{\Phi(s_1) - \Phi(s_2)}{(s_2-s_1) s_2 \Phi(s_2)} \nn
 p(m \ge 1,s_1,s_2) &=& 
 \frac{s_2 \Phi(s_2) - s_1 \Phi(s_1)}{(s_2-s_1) s_2 \Phi(s_2)}
 \Phi(s_1) (1-s_1 \Phi(s_1))^{m-1}
 \eeqa

where $s_1$ is associated to $\tau$ and $s_2$ to $t'$.
In fact one notices that

\beq
 Q(m \ge 1,s_1,s_2) = 
 \left( \frac{1}{s_1 s_2} - p(m=0,s_1,s_2) \right)
 \Phi(s_1) (1-s_1 \Phi(s_1))^{m-1}
 \eeq

and thus the disorder averages factorize:

\beq
 Q(m \ge 1,\tau,t') = (1 - Q(0,\tau,t') ) * \overline{P(m-1,0|\tau)}
\eeq

where the $*$ denotes convolution over the variable $\tau$.
 This could be expected from the markovian and directed nature of
 the walk. The clock is set back to 0 when they exit the trap.
Thus for two times quantities all one needs to know is the 
two time averaged probability of not moving between $t_w$ and 
$t_w + \tau$ plus the averaged probability of diffusing by $m$
during $\tau$. 

Let us first examine the probability of no motion. Explicit laplace
inversion is simple on the asymptotic form for
$\mu <1$ and it gives:

 \beq    \label{exact}
 Q(m=0,\tau \,t') = \frac{\sin(\pi \mu)}{\pi} 
 \int_{0}^{t'/t} dx (1-x)^{-\mu} x^{\mu-1} = F(\tau/t') ~~~ \mu < 1
 \eeq
 
 where:
 
 \beq   
 F(z) = \frac{\sin(\pi \mu)}{\pi} 
 \int_{z}^{\infty} dy \frac{1}{y^{\mu} (1+y)}
 \eeq
 
 One has:

 \beq  \label{assymptot}
F(z) \sim \frac{\sin(\pi \mu)}{\mu \pi} z^{- \mu} ~~~~ z \to \infty
 \eeq

and $1- F(z) \sim z^{1-\mu}$ at small $z$.

It has a aging form as a function of $\tau/t_w$.
This is the manifestation
of the Feigelman Vinokur trap model mechanism of aging. This 
expression is similar to the one obtained in \cite{dean}
in an infinite range model. In the present case
however we are also interested in the diffusion regime
which we now analyze.

We see on the above expression that
if $\tau$ is finite and $t_w \to \infty$
the probability of being trapped is 1. There is no motion on finite
 time scales. We note that this
directed model is not rich enough to contain 
information about the dynamics inside traps. The quasiequilibrium 
regime is thus degenerate $Q(z,\tau) = \delta_{z,0} \delta(\tau)$.

The above result also shows that 
at time $t_w + \tau$ the fraction of particules released by the well 
is of order $(\tau/t_w)^{1-\mu}$ and that 
the particles which are released do fast motion
(as if the clock is then set back to 0 when they exit the trap).
In this model the later motion is not slower.
Thus they will move typically by $\delta x \sim \tau^{\mu}$.

Thus can easily estimate the moments:

\beq
 \overline{ <(x(t_w + \tau ) - x(t_w))^n > }  \sim (\tau/t_w)^{1-\mu} \tau^{n \mu}
 \eeq
 
Note that this gives, for the first moment:

 \beq
 \overline{ <(x(t_w + \tau ) - x(t_w)) > }
 \sim \tau/t_w^{1-\mu}  \sim (t_w + \tau)^{\mu} - t_w^{\mu}
 \eeq

in the regime $\tau << t_w$ consistent with the known result
for the first moment.

We note that there is, in some sense an ``aging regime'' for each 
moment. Indeed one can impose that $\overline{<(x(t) - x(t'))^n>}$ 
is a fixed number while taking both $t_w \to \infty$ and $\tau \to \infty$ 
provided:

 \beq
\tau \sim t_w^{\frac{1- \mu}{1 + (n-1) \mu}} 
 \eeq

This depends on the moment itself. This illustrates the strong non self averaging
properties in one dimension. 

In addition to a self averaging aging regime which is confined to $z=0$
there is in this model a diffusion regime for $z \sim t^\mu$ and
$t \sim t'$. It can be written as:

\beq
\overline{Q(z,t,t')} = \frac{1}{t^{\mu}} F[ \frac{z}{t^{\mu}} , \frac{t'}{t} ]
\eeq

Using the above exact relation:

\beq
\overline{Q(z,t,t')} = \overline{Q(z=0,t,t')} \overline{P(z,t-t',0)}
\eeq

where $P(x,t,0)$ is the single time diffusion front which takes
a scaling form $P(x,t,0) \sim t^{-\mu} \hat{P}(x t^{-\mu})$ 
where $\hat{P}$ was determined in 
\cite{pld_thesis,bouchaud_1d}

One finds:

\beq
F[y,\lambda] =  (1-\lambda)^{-\mu} F[\frac{1-\lambda}{\lambda}] \hat{P}[y (1-\lambda)^{-\mu}]
\eeq

The response function can also be computed. Using the
formula (\ref{formular}) of the Appendix A one sees that the matrix
$B=H$ and thus:

\beq
R(t,t') = \frac{1}{2} \frac{d <x(t)>}{dt}
\eeq
 
and is thus independent of $t'$. In fact any response function
can be obtained that way. For any observable depending only
on time $t$:

\beq
\frac{\delta <O(t)>_h}{\delta h_{t'}} = \frac{1}{2} \frac{d <O(t)>}{dt}
\eeq

which comes from the directed nature of the model.

 \section{general discussion and barrier models} \label{barriers}

In this Section we first discuss aging and diffusion properties 
in low dimensional phase space in terms of scaling of barriers.
To illustrate how these ideas work in practice, we study some cases
which can be solved analytically, such as the directed model of the last Section
with more general wide distributions. These ideas can in principle be applied
to study aging in a wider class of models.

 \subsection{aging, diffusion and scaling of barriers}
 
 One way to formulate the question of aging is to ask
 what is the typical size of the {\it next large barrier}
 typically encountered by the particle {\it after} time $t_w$
(i.e at times $t>t_w$). The idea is that small barriers have already
thermalized. One can usually consider that all barriers of size $T \ln t_w - T C$
 have already thermalized, where $C$ is a constant which can
be chosen large but can be kept fixed as $t_w \to \infty$.
(this will be true in `reasonable' landscapes with fast enough growing
barriers, e.g such that diffusion times are smaller than equilibrium times).
Thus one must look
 at the next barrier encountered after $t_w$ which has a size
 bigger than $E_b  > T \ln t_w - C$.
For a relaxation behaviour of the type $t/t_w$ to occur
 one must have that a typical particle in a typical environment
(i.e that there is a finite fraction of particles and environments)
must overcome a barrier within the range
 $T \ln t_w - T C < E_b < T \ln t_w + T C $ after time $t_w$.
This is illustrated in Fig. 15.
 If typically there is no such barrier, e.g the next encountered
barrier is always larger, then the particle is
 typically thermalized for $t/t_w = c$ and thus one expects:
 
 \begin{eqnarray} \label{degenerate}
 \lim_{t_w \to \infty} \overline{Q(z, c t_w , t_w)} = \overline{Q_{eq}(z,\tau=\infty)}
 \end{eqnarray}
 
 which seems to hold for the Sinai model from our
 simulation results (see section II). This situation is
illustrated in Fig. 16.
 
 On the other hand, if the particles typically encounter such barriers,
then one can have a scaling
 as $t/t_w$. For that one needs an appropriate degeneracy
of the barriers. This is the case for the directed model
 which has en exponential distribution of
 barriers $e^{-a E}$. At $t_w$ there is a finite
 probability $\sim e^{-a ( E_b - T \ln t_w) }$
 that the next barrier is $E_b$. A more precise analysis is
presented in the next section.

\begin{figure}

\label{fig15}

\centerline{\fig{8cm}{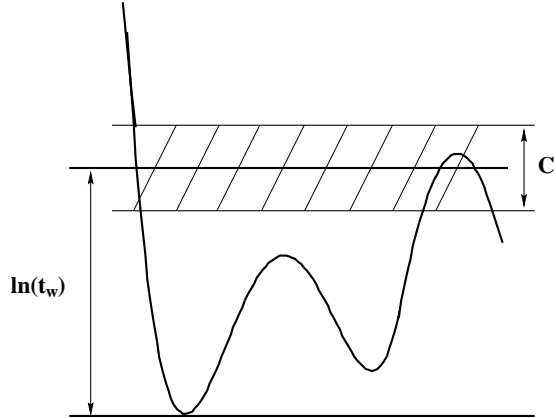}}

\caption{Barrier condition which leads to aging as $t/t_w$}

\end{figure}

 To have aging in $t/t_w$ there is of course no need
 to have a scenario a la Feigelman-Vinokur and also
 no need for randomness. In the next section we
 construct a simple model which exhibits a variety of
 aging and diffusion regimes, {\it but} which satisfies
 the above mentionned property.

\begin{figure}

\label{fig16}

\centerline{\fig{8cm}{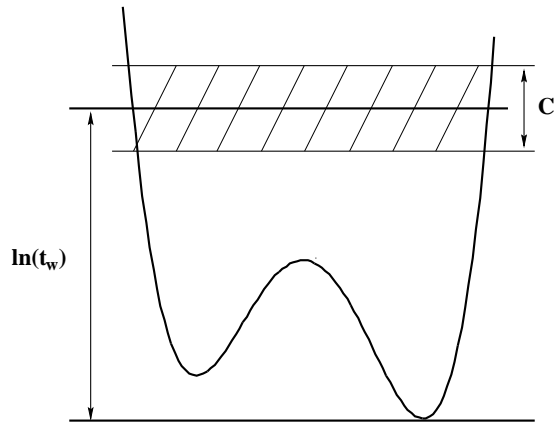}}

\caption{Barrier in Sinai's model}

\end{figure}

 The property (\ref{degenerate}) does not mean that there is
 no aging, but rather that it is degenerate in $t/t_w$.
 In fact (\ref{degenerate}) suggest that there must be
 a scaling which gives some aging. One can always define a function
 $t=g_c(t_w)$ such that:
 
 \begin{eqnarray} \label{def}
 \overline{Q(z=0, g_c(t_w) , t_w)} = c ~~~~~ 0 < c < \overline{Q_{eq}(z=0,\tau=\infty)}
 \end{eqnarray}
 
 The question is how does the function $g_c(t_w)$ behaves for
 large $t_w$. A natural possibility is that $g_c(t_w) \sim t_w^{\mu(c)}$
 and indeed we now argue that this is the case in Sinai's model.
 Note that more generally one can define for each $z$ a function
 $g(c,z,t_w)$ such that $\overline{Q(z=0, g(c,z,t_w) , t_w)} = c$.
If the large $t_w$ behaviour of this function is the same for
all $z$ then one has an aging regime with a non trivial $z$ dependence
as in the models of Section (\ref{solvable}).

 In the symmetric Sinai model the question of the next large barrier
 to be encountered is difficult to answer analytically
 but the simple following arguments can be made.
 Typically at time $t_w$ the particle is within a
 valley (see figure 16). Let us call $U_{min}$ the lowest
 energy point of the valley. The walls of the valley are
 at least $U_{min} + T \ln t_w - T C$. The particle sits 
 at a point which is within $U < U_{min} + T C$ in energy
 (since the valley is typically thermalized, sitting
 any higher has a vanishingly small probability). 
 
 Geometrical construction of possible such valleys in the energy
 random-walk landscape of Sinai's model shows that typically
 the next barrier, i.e the size of the smallest wall
 is of the order of $(1+a) T \ln t_w$ where $a$ is a fluctuating 
positive random variable. This can be seen
 from the fractal nature of the landscape. This suggests
 that a dependence of type $\ln t/\ln t_w$, i.e:
 
 \begin{eqnarray}  \label{sinai}
 \lim_{t,t_w \to \infty} \overline{Q(0, t , t_w)} = F[\frac{\ln t}{\ln t_w}]
 \end{eqnarray}
 
 which is indeed observed. Note that one also expects, from the argument
after (\ref{def}) that an aging scaling form should exist for any finite $z$:

 \begin{eqnarray}
 \lim_{t,t_w \to \infty} \overline{Q(z, t , t_w)} = F[z, \frac{\ln t}{\ln t_w}]
 \end{eqnarray}

Thus to predict aging properties one needs to know the
statistical properties of sequence of barriers effectively encountered
by a particle. It would be appealing to relate it to the
geometry of the environment. A first step would be to 
define in a one dimensional landscape the {\it sequence of next largest barrier}
encountered starting from an initial point.
Let us call them $G_n =E_b(n)$. Two examples of landscapes are
shown in Fig. 17 and Fig. 18. Fig. 17 represents e.g the directed
model of last section, with an arbitrary distribution $P(E)$ of 
barriers (of height $E_i$ corresponding to waiting times $\tau_i=\exp(-E_i/T)$)
which generalizes the algebraic waiting times distribution.
The important question is
how does the series $G_n =E_b(n)$ grow typically with $n$.
This will already give a rough idea of the type of
aging one can expect. In particular it may determine
the aging scaling function $h(t)$. In Sinai model the series
follows a (random) geometric progression $E_b(n) \sim \exp(c n)$ 
as illustrated in Fig. 18 which
leads to the observed behaviour ($\ref{sinai}$).
In order to have a $t/t_w$ behaviour one needs that $E_b(n)$ grows
typically as $c n$ which is what happens in the directed model.
Note that the progression of barriers $E_b(n)$ with $n$ depends only
on the topology of the lansdcape and for instance of distances along
$x$ (different landscapes can have the same $E_b(n)$).

\begin{figure}

\label{fig17}

\centerline{\fig{8cm}{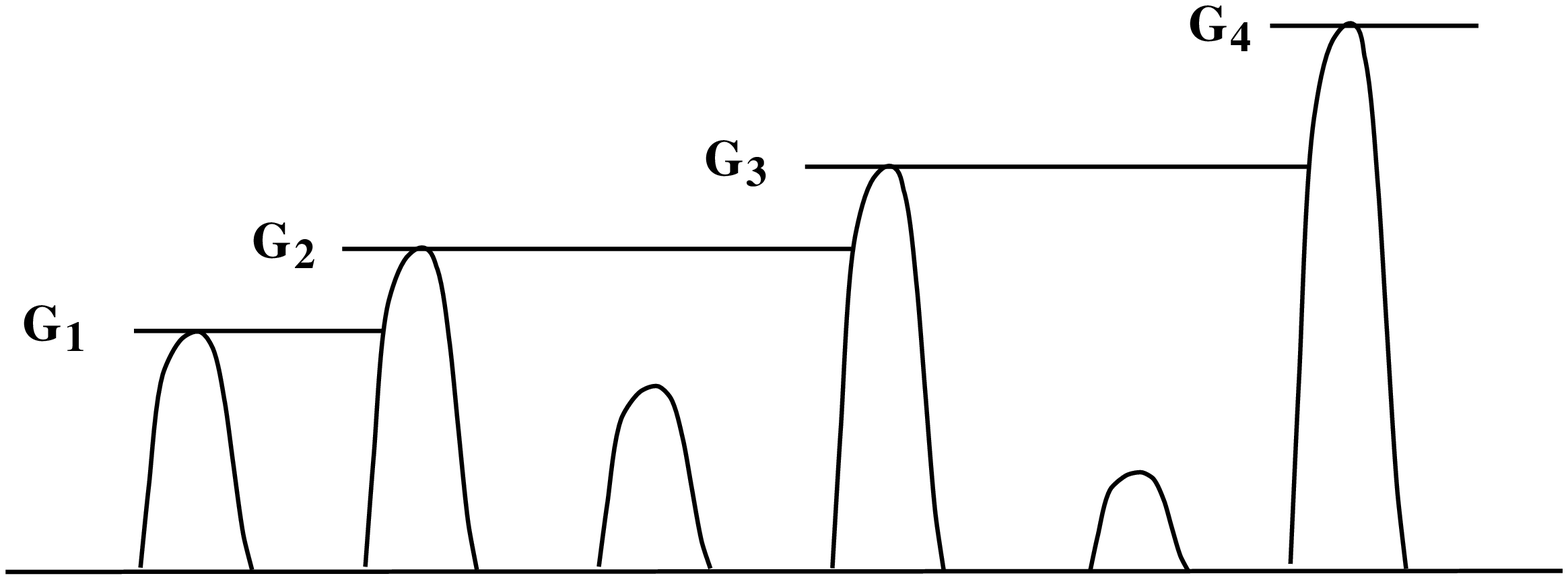}}

\caption{Sequence of next largest barrier in simple models
(see text)}

\end{figure}

\begin{figure}

\label{fig18}

\centerline{\fig{8cm}{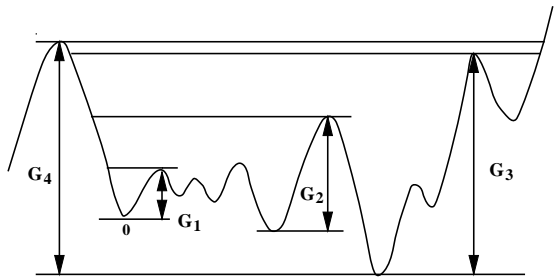}}

\caption{Sequence of next largest barrier in Sinai's model
starting form point $0$}

\end{figure}

Of course to predict more detailed properties one may need to know more: i.e
the sequence of barriers {\it effectively} encountered by the particle.
For this one needs also to know about the well depths. Let us consider e.g.
a non directed model but with a landscape as in Fig. 17 where all the
wells have the same depth. There the packet will
thermalize in a large region (bounded by the next largest barriers)
of size $L(t)$. This will result in a geometrical prefactor
e.g $Q(z=0,t,t') \sim L(t)^{-1} F[h(t)/h(t')]$.
If the well depths grow sufficiently fast then the packet 
will be concentrated at the bottom and then one may expect a
reduced, or even finite prefactor $L(t)=cst$ (as in Sinai's model).

Thus if one knows the statistical properties of 
the {\it sequence of next largest barrier} one knows
a lot about the aging form. In fact if these barriers
grow sufficiently fast with $n$ (faster than $n$) we
expect that this information is enough to determine
these aging form {\it entirely} (provided the wells grow also
fast enough - see remarks of previous section). This suggests,
e.g in Sinai's model, a program to study aging exactly
(though it is technically difficult). Let us illustrate these
consideration on models where this construction can be done
easily.

\medskip

{\it classification of aging forms and constraints from
the CDK bounds}

\medskip

Before we do so let us present some general considerations
about the aging functions $h(t)$. Remember than there is some 
gauge freedom in choosing them since aging quantities are
determined by a fixed ratio $h(t)/h(t_w)=c$. Thus e.g the
choice $h(t) \sim t$ is as good as $h(t)=t^n$
(with the proper change in $c$).

An obvious classification is to distinguish between
three classes of functions $h(t)$.

(i) $h(t)$ grows slower than t (or any power of $t$).
This corresponds to fast growing barriers. The condition
$h(t)/h(t_w)=c$ can also be expressed as
$\tau=t-t_w \sim t \sim H(c,t_w)$. Examples are:

\beqa
&& h(t) \sim \ln t ~~~~  t \sim \tau \sim t_w^c ~~~ (c>1) \\
&& h(t) \sim e^{ (\ln t)^a } ~~~~~
t \sim \tau \sim t_w \exp( \frac{\ln c}{a} (\ln t_w)^{1-a}) ~~~ 0<a<1 \\
&& h(t) \sim e^{ (\ln t)/(\ln \ln t) } ~~~~~ t \sim \tau \sim t_w (\ln t_w)^{\ln c}
\eeqa

(ii) simplest aging behaviour $h(t)=t$.

(iii) $h(t)$ grows faster than t (subaging). Then one must have
$\tau=t-t_w \ll t_w$, such as in the following examples:

\beqa
&& h(t) \sim e^{ (\ln t)^a }  ~~~~  \tau= \frac{\ln c}{a} \frac{t_w}{(\ln t_w)^{a-1}}
~~~ a>1 \\
&& h(t) \sim e^{ t^a } ~~~~~ \tau \sim \frac{\ln c}{a} t_w^{1-a} ~~~ 0<a<1 \\
\eeqa

The CDK bounds discussed in Section II also put general
constraints on possible aging forms assuming they have, as in mean field,
a non trivial $0< X \neq 1 < \infty$ (see Appendix B for definitions).
Assume indeed that as in mean field all correlation functions
and $X$ are functions of $h(t')/h(t)$. Then bound 
(\ref{bound2}) implies that:

\beqa
\frac{d \ln h(t')}{dt'} \ll |\frac{d \overline{H}(t')}{d t'}|^{1/2}
\eeqa

Thus if $\overline{H}(t') \sim {t'}^{-\alpha}$ one has that

\beqa
h(t) \ll \exp( t^{(1-\alpha)/2} )
\eeqa

which excludes a large class of subaging behaviours (Sinai corresponds
to $\alpha=0$).

\subsection{sequence of largest barriers}

It is simple in some cases to determine exactly the distribution of the
sequence of next largest barriers.

Let us look again at the directed model defined by
a set of successive barriers $E_i$, identically distributed
with a distribution $P(E)$.
We denote $H(E) = \int_E^{+\infty} P(E') dE' = \text{Proba}(E'>E)$

Let us estimate the probability density that the sequence of successive
next largest barriers (see Fig. 17) be
$E_0,G_1,G_2,..G_n$. It is by definition:

\beqa
&& Q(E_0,G_1,..G_n) = \sum_{k_1=0, \infty} ..\sum_{k_n=0, \infty} 
\text{Proba}(E_1<E_0,..E_{k_1}<E_0,E_{k_1+1}=G_2>E_0,E_{k_1+2}<G_1,..E_{k_2+1}=G_2>G_1,...\\
&& .. E_{k_n+1}=G_n>G_{n-1})
\eeqa

This yields immediately:

\beqa  \label{hello}
Q(E_0,..G_n) = p(E_0) \frac{p(G_1) \theta(G_1-E_0)}{H(E_0)} 
\frac{p(G_2) \theta(G_2-G_1)}{H(G_1)}...
\frac{p(G_n) \theta(G_n-G_{n-1})}{H(G_{n-1})}
\eeqa

It is easy to see that the exponential distribution
$p(E) = \exp(-E) \theta(E)$ has a special property. 
Indeed, in that case:

\beqa
Q(E_0,..G_n) = \theta(E_0) \theta(G_1-E_0) \theta(G_2-G_1) ..\theta(G_n-G_{n-1})
e^{-G_n} 
\eeqa

but this can be written as the product:

\beqa
Q(E_0,..G_n) = \prod_{i=0,1,..n} p(G_i-G_{i-1})
\eeqa

with $G_0=E_0$ and $G_{-1}=0$. This shows that the difference between
successive largest barriers have {\it the same exponential distribution}
$p(E)$ and are {\it independently distributed}. Thus:

\beqa
G_n = \sum_{i=0,n} w_i 
\eeqa

The central limit theorem can then be used and leads to a Gaussian
distribution for the variable $(G_n - n)/\sqrt{n}$.
This remarkable property of the exponential distribution allows to
understand the Feigelman Vinokur scenario for aging as a 
linear (random) growth of the next largest barrier $E_b(n)=G_n
\sim c n$

This remarkable property of the exponential distribution also
allows to obtain the general solution of the problem of
finding the probability of a sequence of largest barrier.
Defining $\Phi(E) = - \ln \int_E^{+\infty} dE' P(E')$ one has:

\beqa
P(E) dE = d e^{- \Phi(E)} \theta(E) = \Phi'(E) e^{- \Phi(E)} \theta(E) dE
\eeqa

and thus (\ref{hello}) can be put in the form:

\beqa
&& Q(G_0,..G_n) dG_0..dG_n = \\ d\Phi(G_0)...d\Phi(G_n) 
&& \theta(G_0) \theta(G_1-G_0) \theta(G_2-G_1) ..\theta(G_n-G_{n-1})
e^{-( (G_n-G_{n-1}) + (G_{n-1}-G_{n-2}) +..+ (G_1-G_0) + G_0)  } 
\eeqa

and the sequence of variables $\Phi(G_0),..\Phi(G_n)$
can be constructed as a ``random walk'':

\beqa
\Phi(G_n) = \sum_{i=0}^n w_i
\eeqa

where the $w_i$ are a set of independent variables, identically distributed
with an exponential distribution $P(w) dw = e^{-w} \theta(w) dw$.
There are several consequences. First
the distribution of each $\Phi(G_n)$ is thus a Poisson process:

\beqa \label{distrib}
Q(G_n) d\Phi(G_n) = \frac{\Phi(G_n)^{n}}{n!} e^{-\Phi(G_n)}
\eeqa

and one has the central limit theorem for large $n$:

\beqa
\frac{\Phi(G_{n}) - n}{\sqrt{n}} \to v
\eeqa

where $v$ is a centered gaussian random variable of unit variance.

One can apply these results to various cases:

(i) algebraically growing barriers:

$P(E) dE = a E^{a-1} e^{-E^a} \theta(E) dE$ which corresponds to 
$\Phi(E) = E^a$.  Then one has:

\beqa
G_n \sim n^{1/a} 
\eeqa

Note that the randomness (disorder) is only a 
subleading correction.

(ii) exponentially growing barriers:

$P(E) dE = (1+E)^{-(a+1)} \theta(E) dE$ which corresponds to 
$\Phi(E) = a \ln E$.  Then one has:

\beqa
G_n \sim e^{n/a}
\eeqa

The variable $G_n$ has a log normal distribution.

\subsection{consequences: aging properties}

Having determined exactly the sequence of next largest barriers
one can construct a toy model which will mimic the exact 
diffusion process. We claim that if barriers grow fast enough
(faster than $E_n \sim n$) it will give the exact aging behaviour.
It can be applied to a variety of landscape, but let us apply
it here only to the previously considered directed model.

The toy model amounts to approximate the quantity $\overline{Q(z,\tau,t_w)}$
at $z=0$ (i.e the probability to remain still between $t_w$ and
$t_w + \tau$) as:

\beqa
\overline{Q(\tau,t_w)} \equiv \overline{Q(0,t_w + \tau,t_w)} = \sum_{k=0}^{+\infty}
\theta(G_{k-1} < T \ln t_w < G_{k}) \exp(- \tau e^{-G_k/T})
\eeqa

This means that at $t_w$ all barriers smaller than $T \ln t_w$ have been
overcome and that the only relaxation process is to go over the
next highest barrier. This is illustrated in Fig. 17
which is adequate for the directed model.
It thus supposes that the probability to be in another
well is zero, which is correct for the directed model if barriers grow 
fast enough (which means distribution of waiting times wider than power laws)

This can be rewritten as:

\beqa
Q(\tau,t_w) = \sum_{k=0}^{+\infty}
(\theta(\Phi_w - \Phi_{k-1})  - \theta(\Phi_w - \Phi_{k}))
\exp(- \tau e^{- \frac{1}{T} \Phi^{-1}(\Phi_k)})
\eeqa

where $\Phi_k = \Phi(G_k)$ and $\Phi_w = \Phi(T \ln t_w)$.
Because of the above statistical
properties of the sequence this yields upon averaging:

\beqa
\overline{Q(\tau,t_w)} = \int d\phi 
\sum_{k=0}^{+\infty} Q_k(\phi)
\int_0^{+\infty} dw e^{-w}
(\theta(\Phi_w - \phi + w)  - \theta(\Phi_w - \phi))
\exp(- \tau e^{- \frac{1}{T} \Phi^{-1}(\phi)})
\eeqa

and we can use (\ref{distrib}) namely that 
$\sum_{k=0}^{+\infty} Q_k(\phi)=1$.
In the large time limit one can shift the 
integrand globally by $\phi \to \phi + \Phi_w$ without
edge effects (though one must
be careful not to shift terms independently because of
divergent integrals). Thus:

\beqa
\overline{Q(\tau,t_w)} = \int d\phi \int_0^{+\infty} dw e^{-w}
(\theta(- \phi + w)  - \theta(- \phi))
\exp(- \tau e^{- \frac{1}{T} \Phi^{-1}(\phi + \Phi_w)})
\eeqa

This can be rewritten, after integration by parts as:

\beqa \label{resultgeneral}
\overline{Q(\tau,t_w)} = \int_0^{+\infty} dv e^{-v}
\exp(- \tau e^{- \frac{1}{T} \Phi^{-1}(v+ \Phi_w)})
\eeqa

This is our general result for this toy model. 

Let us estimate (\ref{resultgeneral})  first
for models with fast enough growing barriers, i.e faster
than $\Phi(E) \sim E$. We introduce the aging scaling function
$h(t) = e^{\Phi (T \ln t)}$ and thus
$\Phi(x) = \ln h(e^{x/T})$ and $\Phi^{-1}(y) = T \ln h^{-1}(e^y)$.
One can rewrite:

\beqa
\overline{Q(\tau,t_w)} = \int_0^{+\infty} dw e^{-w}
\exp(- e^{- \frac{1}{T} S(w,\tau,t_w)} )
\eeqa

with $S(w,\tau,t_w) = \Phi^{-1}(w+ \Phi(T \ln t_w)) - T \ln \tau$.
It turns out that if barriers grow fast enough the
function $\exp(- e^{- S})$ acts exactly as a theta function $\theta(S)$
and the result is simply:

\beqa \label{result3}
\overline{Q(\tau=t-t_w,t_w)} = \frac{h(t_w)}{h(\tau)} ~~~~ (\tau \gg t_w)
\eeqa

This can be shown by a careful examination of the asymptotics.
The point is that if one takes $\tau$ and $t_w$ to $+\infty$
such that the ratio $h(t_w)/h(\tau)$ is fixed 
then the result is (\ref{result3}). Since barriers grow fast enough
one has $\lim h(t_w)/h(t) = \lim h(t_w)/h(\tau)$. A more correct
version of the above statement is that:

\beqa \label{result30}
\lim_{t_w \to \infty, t \to \infty, \frac{h(t_w)}{h(t)} = y}
\overline{Q(\tau=t-t_w,t_w)} = y
\eeqa

It is interesting to note that one has {\it exactly}:

\beqa  \label{strikingly}
e^w = \frac{h(t)}{h(t_w)}
\eeqa

and thus the above result follows from the exponential
distribution of $w$. Note that this is like taking the $T=0$
limit and thus if barriers grow fast enough we are dealing with a $T=0$ fixed point.
Note that our result can also be rewritten
quite generally as:

\beqa \label{result31}
\overline{Q(\tau=t-t_w,t_w)} =
\frac{\int_{t}^{+\infty} P(\tau) d\tau }{\int_{t_w}^{+\infty} P(\tau) d\tau }
\eeqa

as a function of the waiting time distribution $\tau=e^{E/T}$. In particular
we have determined that the aging function is exactly:

\beqa \label{result32}
h(t) =
\frac{1}{\int_{t}^{+\infty} P(\tau) d\tau }
\eeqa

One can also define the probability that
at time $t_w$ the particle is next to a barrier
(i.e in a site) of waiting time
$\tilde{\tau}$, i.e:
 
\beqa
&& \overline{Q(\tau,t_w)}=
\int_0^{+\infty} d\tilde{\tau} P_{t_w}(\tilde{\tau}) e^{- \tau/ \tilde{\tau}} \\
\eeqa

Then the above formulae yields that this 
``aged'' waiting time distribution is:

\beqa
P_{t_w}(\tilde{\tau}) d\tilde{\tau} = \frac{h(t_w)}{h(\tilde{\tau})}
\frac{d h(\tilde{\tau})}{h(\tilde{\tau})} \theta(h(\tilde{\tau}) - h(t_w))
\eeqa

This also takes the simple form:

\beqa   \label{simplef}
P_{t_w}(\tilde{\tau}) d\tilde{\tau} =
\frac{P(\tilde{\tau}) d\tilde{\tau}}{\int_{t_w}^{+\infty} P(\tilde{\tau '}) d\tilde{\tau '}}
 \theta(\tilde{\tau} - t_w)
\eeqa

which we have shown holds exactly.

We can now apply these results it to several cases.

(i) algebraically growing barriers:

$\Phi(E) = E^a$ with $a<1$. Then one has $h(t) = \exp( (T \ln t)^a )$.

\beqa \label{result4}
\overline{Q(\tau=t-t_w,t_w)} = \exp( - [ (T \ln t)^a - (T \ln t_w)^a ])
\eeqa

(ii) exponentially growing barriers:

$P(E) dE = (1+E)^{-(a+1)} \theta(E) dE$ which corresponds to 
$\Phi(E) = a \ln E$ and $h(t) = (T \ln t)^a$. Then one has:

\beqa
\overline{Q(\tau=t-t_w,t_w)} = (\frac{\ln t_w}{\ln t})^a
\eeqa

Let us now apply the above model in the case of linearly growing
barriers ($\Phi(E)=E$). There we know that the assumption that
the whole packet is concentrated nest to the next highest barrier
is certainly not exact. Indeed since there are many barriers
within $T \ln t_w -C < E< T \ln t_w +C$ the packet will be spread
out (barrier degeneracy). It is instructive though, to see how well
this toy model does in that case.

Setting $y=\tau/t_w$ one gets:

\beqa
\overline{Q(y)} = \int_0^{+\infty} dw e^{-w}
\int_0^w d\phi 
\exp(- y e^{- \phi/T})
\eeqa

After integration by part and change of variables:

\beqa  \label{above}
\overline{Q(y)} = \int_0^{+\infty} dw e^{-w} \exp(- y e^{- w/T})
= T y^{-T} \int_0^y \frac{d \lambda}{\lambda} \lambda^T e^{-\lambda}
\eeqa

For large $y$ it behaves as:

\beqa
\overline{Q(y)} \sim T \Gamma[T] y^{-T}
\eeqa

which is to be compared with (\ref{assymptot}) ($T$ is exactly $\mu$, $T=\mu$, from the
relation between waiting times and barrier heights). The toy model gives exactly
the exponent, but not the prefactor. We note however that 
at small $T$ the prefactor becomes exact which is in agreement with
the fact that for faster growing barriers the toy model becomes 
exact. At small $y$ the toy model however does not 
 yield the non analyticity at $\tau << t_w$, which thus entirely originates
from the initial (fractal) dispersion of the packet at $t_w$ over several wells.

This confirms our physical picture: only maxima start playing a role 
when barriers $G_n$ grow faster than $n$. The case $G_n \sim n$ is the
marginal case when finer information is important.

Note that the exact result (\ref{exact}) can be put under the form:

\beqa
&& \overline{Q(\tau,t_w)}=
\int_0^{+\infty} du \phi(u) e^{- u \tau/t_w} \\
&& \phi(u) = \frac{\sin (\pi T)}{\pi} \frac{e^{-u}}{u} \int_0^u \frac{dt}{t \Gamma[T]} t^T e^t
\eeqa

to be compared with the toy model which has a $\phi(u)=T u^{T-1} \theta(0<u<1)$.
Thus one can see that
for large $w$ the $e^{-w}$ is correct, however the barriers
which are not large, or the one at $w<0$ (smaller than $t_w$)
start playing an important role.

The role of smaller barriers can be illustrated as follows.
The correct distribution of the waiting time $\tilde{\tau}$ of the site where the
particle is at $t_w$ is computed in the Appendix. It reads:

\beqa
P_{t_w}(\tilde{\tau}) d\tilde{\tau} =  \frac{\sin (\pi \mu)}{\pi} \frac{d\tilde{\tau}}{t_w} 
 (\frac{t_w}{\tilde{\tau}})^{1+\mu}
\int_0^{1} e^{- \frac{t_w}{\tilde{\tau}} (1- u) } \frac{u^{\mu-1}}{\Gamma[\mu]}
\eeqa

This should be compared with what the barrier model which gives:

\beqa
P_{t_w}(\tilde{\tau}) d\tau = \mu 
\frac{d\tilde{\tau}}{t_w}  (\frac{t_w}{\tilde{\tau}})^{1+\mu}
\theta(\tilde{\tau}-t_w)
\eeqa

from the above fomulae (\ref{simplef}) and the identification $\tilde{\tau}=e^{w/T}$.
As above, the toy model gives the correct power law dependence for {\it large}
waiting times (the prefactor itself becoming exact when $\mu \to 0$). There
is however an accumulation of smaller
barriers effectively seen by the particle, which is not captured.

Let us conclude this section by noting that this type of distribution 
of next largest barriers can be used to analyze a large variety of models.
For instance the analysis of the symmetric waiting time model in $d=1$ or of directed
model with several branches will be quite similar but goes beyond this paper.

\section{A solvable model with aging and diffusion} \label{solvable}
 
We will now present a solvable model which exhibits simultaneously
a non trivial aging regime and a non trivial diffusion regime. 
The third regime (the quasi-equilibrium one) is degenerate
(it is reduced to a point). More properly this is
really a class of solvable models, and we will only study 
a few.

In view of the discussion about barriers of the preceding
section the best way to construct a one dimensional
diffusion model with a non trivial aging regime (i.e such that
there is a finite probability that $z=x(t)-x(t_w)$ remains
finite when both $t$ and $t_w$ are large) is to make sure that
the next largest barrier seen after time $t_w$ is equal to 
the previous one plus a constant (for aging as $t/t_w$). 
A natural landscape is thus to look at a succession of barriers $E_b^n \sim n$.
However one wants the valleys also to become deeper and deeper,
otherwise the thermal packet will be too extended. Thus a 
natural choice is also to suppose valleys to scale as
$E_{min}^n \sim - n$, a landscape represented in Fig. 19
It turns out that the {\it continuous} version of this model, as
well as some generalizations, can be solved exactly in a very simple way.

\begin{figure}

\label{fig19}

\centerline{\fig{8cm}{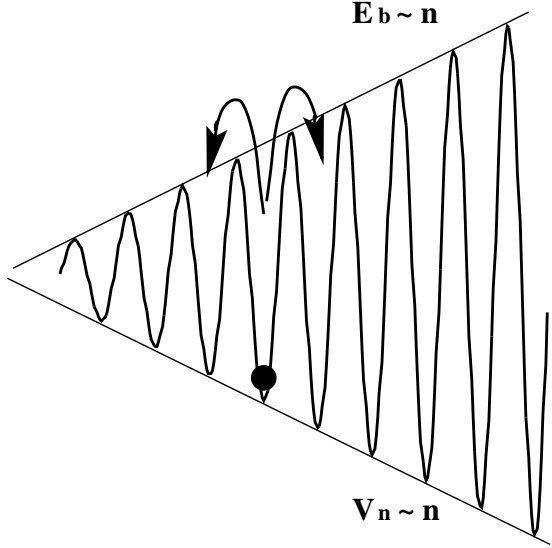}}

\caption{Aging model studied in the text in the continuum limit}

\end{figure}

\subsection{the general model and its solution}

 Let us consider the following one dimensional diffusion equation:
 
 \begin{eqnarray}  \label{diffusion}
 \partial_t P(x,t) = \partial_x ( D(x) ( \partial_x P(x,t) - F(x) P(x,t) ) )
 \end{eqnarray}

This model corresponds to diffusion in a landscape with both
barriers $E_b(x)$ and a potential $U(x)$ (valleys) such that:
 
 \begin{eqnarray}
 D(x) = e^{- E_b(x) } ~~~ U'(x) = - F(x)
 \end{eqnarray}
 
here and in what follows we will set the temperature $T=1$ for
convenience but it can be easily put back in. The potential
is defined by the fact that the equilibrium zero current measure
is $P_{eq}(x) = \exp(-U(x))$. 
The general model (\ref{diffusion}) cannot be solved but
there is a particular case which can be solved easily, which is:

 \begin{eqnarray}
U(x) = - E_b(x)/2
 \end{eqnarray}

The model is defined by giving a function $\Phi(x)$ such that:

 \begin{eqnarray}
e^{U(x)} = e^{- E_b(x)/2 } = \Phi(x)
 \end{eqnarray}

Then the diffusion equation becomes:
 
 \begin{eqnarray}
 \partial_t P(x,t) =
 \partial_x ( \Phi(x) \partial_x ( \Phi(x) P(x,t) ) )
 \end{eqnarray}
 
and is easily solved since one can define a new variable
$u$ and a new probability $G(u,t)$
 such that:
 
 \begin{eqnarray}
 \frac{du}{dx} = \frac{1}{\Phi(x)} ~~~ 
 P(x,t) dx = G(u(x),t) du(x)
 \end{eqnarray}
 
 In the variable $u$ the system is a free
 diffusion problem !
 
 \begin{eqnarray}
 \partial_t G(u,t) = \partial^2_u G(u,t)
 \end{eqnarray}
 
 Clearly for the choice $\Phi(x)=e^x$ and thus
 $E_b(x) = 2 x$ this model is some continuum limit of
 the one represented in Fig. 19. It will have the
same expected large time properties. Note that the quasi-equilibrium 
regime will be lost in this limit (i.e reduced to
a delta function, see below) as the size of each well 
will become infinitesimal. It would be nice to
be able to solve directly the model of Fig.19.
to obtain also the small times behaviour.
In any case the results presented here for large times
will be the same (they are not an artefact
of the continuous limit).

One must define carefully the boundary conditions. 
There are basically two choices that we will study:

{\it free boundary conditions}

One can study a barrier landscape $E_b(x)$ defined from
 $- \infty < x < \infty$. This could either be a random landscape
$U(x)$ or a deterministic one with growing barriers
in $d=1$ which we then should choose symmetric for definiteness
for instance $U(x) = - |x|$. Then a natural definition 
 of $u$ is :
 
 \begin{eqnarray}  \label{ux}
 u(x) = \int_0^x dx' e^{- U(x')}
 \end{eqnarray}
 
 and $- \infty < u < \infty$.
 
 The Green function is thus simply unbounded diffusion
 
 \begin{eqnarray}
 G(u,t| u_0,0) = \frac{1}{\sqrt{4 \pi t}} e^{-\frac{(u-u_0)^2}{4 t} } 
 \end{eqnarray}
 
 which yields the Green function for the original equation:
 
 \begin{eqnarray}
 P(x,t| x_0,0) = \frac{1}{\sqrt{4 \pi t}}
 e^{- U(x)}  e^{- \frac{ (\int_{x_0}^x  dx' e^{- U(x') } )^2}{4 t} } 
 \end{eqnarray}
 
 We will be interested in the probability of displacements $z$ between $t'$ and $t$:
 
 \begin{eqnarray}
 Q(z,t,t') = \int dx P(x+z,t|x,t') P(x,t'|x_0=0,0)
 \end{eqnarray}
 
 which formally reads:
 
 \begin{eqnarray}
 Q(z,t,t') = \int dx \frac{1}{4 \pi \sqrt{t'(t-t')}}
 e^{- U(x) - U(x+z)} 
 e^{- \frac{ (\int_{0}^x  dx' e^{- U(x') } )^2}{4 t'}
 -  \frac{ (\int_{x}^{x+z}  dx' e^{- U(x') } )^2}{4 (t-t')} }
 \end{eqnarray}
 
 Another convenient form is to write:
 
 \begin{eqnarray}
 Q(z,t,t') = \int du_1 du_2
 \frac{1}{4 \pi \sqrt{t'(t-t')}}
 \delta(z - (x(u_1 + u_2) - x(u_1)))
 e^{- \frac{ u_1^2}{4 t'}
 -  \frac{ u_2^2}{4 (t-t')} }
 \end{eqnarray}

where $x(u)$ is the function implicitly defined by (\ref{ux}).

{\it reflecting boundary at $u=0$}

Since we will be interested in landscapes with growing barriers
such as depicted in Fig. 19. it is useful in certain cases 
to introduce a reflecting boundary on the left. One can choose
for definiteness a landscape such that $U(x \to - \infty) \to +\infty$,
define $u(x) = \int_{-\infty}^x dx' e^{- U(x')}$ and use a
reflecting boundary at $u=0$. This naturally avoids the particule being either in one
half space or the other. For instance for the lanscape $E_b(x) = 2 x$,
for which $u(x)=e^x$ the reflecting boundary is at $x=-\infty$.
Because of reflecting boundaries one must choose
the free propagator with $\partial_u G(u=0) =0$, i.e:
 
 \begin{eqnarray}
 G(u,t|u',t') = \theta(u) \frac{1}{\sqrt{4 \pi (t-t')}} 
 ( e^{-\frac{(u-u')^2}{4 (t-t')} } + e^{-\frac{(u+u')^2}{4 (t-t')} } )
 \end{eqnarray}
 
We will be interested in:

 \begin{eqnarray}
 Q(z,t,t') = \int dx P(x+z,t|x,t') P(x,t'|x_0= -\infty,0)
 \end{eqnarray}
 
where we have chosen for convenience
the initial condition at $x_0 = - \infty$ (and thus $u_0=0$).
This is a purely technical point on the definition of the
model and has no bearing on the physics. Indeed since we have chosen
$E_b(x \to - \infty) \to -\infty$ and the initial condition will
be immaterial since it takes only a finite time in this
 model to reach finite $x$ values and we are interested only
 in the later (long time) behaviour. Thus using:

 Using $P(x,t|x',t') = \Phi(x)^{-1} G(u(x) , t| u(x') , t')$
 one finds:

 \begin{eqnarray}
 Q(z,t,t') = \int dx \frac{1}{4 \pi \sqrt{t'(t-t')}}
 e^{- U(x) - U(x+z)} 
 e^{- \frac{ (u(x))^2}{4 t'}} ( e^{- \frac{ (u(x+z)-u(x))^2}{4 (t-t')}}
 + e^{- \frac{ (u(x+z)+u(x))^2}{4 (t-t')}} )
 \end{eqnarray}

 Let us conclude this section by indicating that
a more general case can
 be solved (see also \ref{appendixd}), i.e  brought back to:
 
 \begin{eqnarray} \label{dis}
 \partial_t G(u,t) = \partial^2_u G(u,t) - v \partial_u G(u,t)
 \end{eqnarray}
 
 which corresponds to:
 
 \begin{eqnarray}
 U(x) = - ( E_b(x)/2 + v \int_0^x dy e^{E_b(y)/2} )
 \end{eqnarray}
 
These models corresponds to either valleys and barriers
scaling differently and we will not study these models here.
We note that the directed model of section \ref{directed}
can be seen as a particular case of the class of models
introduced here. Indeed the case (\ref{dis}) corresponds
to the diffusion in $x$ given by:

 \begin{eqnarray}
v t' + \sqrt{t'} w  = u(x)- u(x_0) = \int_{x_0}^{x} e^{- U(x')} dx'
 \end{eqnarray}

where $w$ is a normalized gaussian variable. In the fully 
directed case (large $v$) and for a judicious choice of
the (random) $U(x)$ one can recover the directed models.

 \subsection{solution for linearly growing barriers}
 
 It corresponds to the model of Fig. 15. Again one can either 
consider the symmetric
landscape taking $U(x) = - |x|$ and thus $x(u) = sgn(u) \ln(1 + |u|)$.
Or one can take the
``half landscape'' with a reflexive barrier at $x_0 = -\infty$
which we will discuss first.

 The above formulae give in that case, for the single
time packet:

\begin{eqnarray}
 P(x,t| x_0,0) = \frac{2}{\sqrt{4 \pi t}}
 e^{x}  e^{- \frac{ e^{2 x}}{4 t} } 
 \end{eqnarray}

The generating function of moments is $<e^{ \lambda x}> = \Gamma[(1+\lambda)/2]
(4 t)^{\lambda)/2}/\sqrt{\pi}$. Thus one has:

\begin{eqnarray}
&& <x(t)> = \frac{1}{2} \ln(t/c) \\
&& <x^2(t)> -<x(t)>^2 = \pi^2/8
\end{eqnarray}

where $c=\ln \gamma$ ($\gamma$ is Euler's constant). In fact
the packet has a limit shape as one can see by performing the
shift $x = \frac{1}{2} \ln(4 t) + \tilde{x}$. Then the distribution
of $\tilde{x}$ is asymptotically time independent:

\begin{eqnarray}
 P(\tilde{x},t) \sim \frac{2}{\sqrt{\pi}}
 e^{x}  e^{- e^{2 \tilde{x}}} 
 \end{eqnarray}

Thus the packet has a constant width and simply spreads over
a few wells (in Fig. 15) with its center moving logarithmically
towards the right. Thus, since there is some degeneracy of barriers
one expects a $t/t_w$ aging behaviour.

Indeed one gets for the two time packet:

 \begin{eqnarray}
 && Q(z,t,t') = \int_0^{+ \infty} du u
 \frac{e^z}{2 \pi \sqrt{t' (t-t')}} e^{- \frac{u^2}{4 t'} }
 (
 e^{-\frac{u^2 (e^z-1)^2}{4 (t-t')} } + e^{-\frac{u^2 (e^z+1)^2}{4 (t-t')} }
 )
 \end{eqnarray}
 
This yields to an aging form for the distribution of displacements
between $t'$ and $t$, given by:

\begin{eqnarray}  \label{resultaging}
 && Q(z,t,t') = \frac{e^z}{\pi} (
 \frac{ \Gamma }{ (e^z -1)^2 + \Gamma^2 }
 + \frac{ \Gamma }{ (e^z + 1)^2 + \Gamma^2 } ) ~~~ \Gamma=\frac{\sqrt{t-t'}}{\sqrt{t'}}
 \end{eqnarray}
 
which, in the variable $w = e^z$, is the sum of two Lorentzian of width $\Gamma$
the first one centered around $w=1$ and the second one which is its mirror image
(with the mirror at $w=0$). This result is natural considering that
the {\it ratio} of two independent gaussian variables 
with unit variance $v=v_1/v_2$ is the Lorentzian $P(v)=1/(\pi(v^2+1))$
and that one has:

\begin{eqnarray}
z = x(u_1 + u_2) - x(u_1) = \ln (1 + \frac{u_2}{u_1}) = \ln (1 + \Gamma v)
\end{eqnarray}

where $u_1 = \sqrt{t'} v_1$ is the positive gaussian variable representing
the diffusion process on the 
half line and $u_2 = \sqrt{\tau} v_2$ another gaussian variable representing
the later diffusion process, constrained so that the sum $u_1 + u_2$ remains
positive, hence the mirror. The total weight in the second packet is $p=(\arctan \Gamma)/\pi$
and in $1-p$ the first.

Thus the first Lorentzian packet in (\ref{resultaging}) corresponds to particles
which have remained in the region $x(t')$ while the others have crossed to the mirror
at $x=-\infty$ at least once and came back to that region. We are mostly interested
with this first packet (which always contains a fraction $>1/2$ of particles)
but the second packet will always be there as a mainly technical feature of
the model. If we chose instead the symmetric environment, these would be two 
separate aging packets one around $x(t')$ and the other around $-x(t')$. Aging
then occurs only within each packet (there are diffusion events between packets)
while here one has aging in all the packet. 

Indeed the above distribution (\ref{resultaging})
clearly exhibits an aging scaling form of the type:
 
 \begin{eqnarray}
 Q(z,t,t') = \tilde{Q}(z, \frac{t}{t'} ) = \tilde{Q}(z, \frac{h(t)}{h(t')} )
 \end{eqnarray}
 
 thus $h(t) = t$ in this model. The function $\tilde{Q}$ indeed depends on the times
only through $\Gamma$ which itself can be written as:

 \begin{eqnarray}
 \Gamma = f[\frac{h(t)}{h(t')}] 
 \end{eqnarray}
 
 where the form of $f(x)$ and $h(t)$ is univoquely determined as
 $h(t) = t$ and $f(x) = \sqrt{1-x}$. Thus, as in mean field \cite{Cule} there is
a {\it singularity} at the beginning of the aging regime: the $\beta$
 exponent is equal to $\beta = 1/2$.

It is interesting to note though that at the beginning of the
 aging regime $\Gamma << 1$, there is anomalous behaviour
 of the moments $z^n$, because the Lorentzian has diverging moments:
 
 \begin{eqnarray}
<z^n> =  \int_{-1/\Gamma}^{+ \infty} (\ln(1 + \Gamma w) )^n \frac{dw }{1+ w^2} 
 \end{eqnarray}

One finds in particular $<z> = 1/2 \ln(1 + \Gamma^2) = 1/2 \ln(t/t')$ exactly.
 
 \begin{eqnarray}
 <(z - <z>)^n> = \frac{\Gamma}{2 \pi \sqrt{1 + \Gamma^2} }
 \int_{-\infty}^{+\infty} du u^n  ( \frac{1}{\cosh[u] - \frac{1}{\sqrt{1 + \Gamma^2}} }
 + \frac{1}{\cosh[u] + \frac{1}{\sqrt{1 + \Gamma^2}} } )
 \end{eqnarray}
 
 One finds that for $n>1$ $<(z - <z>)^n>  \sim \Gamma \Gamma[n+1]$
 (i.e $P(u) = \Gamma e^{-u}$) which is strong intermittence.
 
 For widely separated time scales there is also a {\it diffusion regime}
in this model. At large $t$ one gets:
 
 \begin{eqnarray}
 z = \ln(v) + \ln(t) - \ln(t')
 \end{eqnarray}
 
where $v$ has a Lorentzian distribution. Thus the diffusing packet has
a {\it finite} size (note that since for a Lorentzian $<\ln(v)>  = 0$
(inversion symmetry) one recovers the above result). Thus there is
a diffusion regime defined by $z,t,t' \to \infty$ with
$z/\ln t$ and $\ln t/ \ln t'$ fixed. It has no thermal fluctuations (it is completely
determinist) and the diffusion scaling function reads:
 
 \begin{eqnarray}
Q(z,t,t') dz \sim \frac{dz}{\ln t}
\delta( \frac{z}{\ln t} - (1 - \frac{\ln(t)}{\ln(t')}) )
 \end{eqnarray}
 
 As in SR mean field models the aging regime smoothly merges in
 the diffusion regime \cite{Cule}

 This model thus contains both an aging regime in $t/t_w$
 and a diffusion regime with a different scaling. 
 Only the FDT regime cannot be seen, since it has
 disappeared in the continuum limit.
 
Finally note that this model seems to 
violate the quasistatic assumption. Indeed in a box of finite size $L$,
or with periodic boundary conditions
and taking $t_w \to \infty$, the packet will converge towards the
equilibrium measure. Since $U(x) = -x$ this will result in an
equilibrium packet of {\it finite size}. However the 
dynamical regime $t-t' \ll t'$ leads only to a delta funtion
packet at $\delta(z)$, thus totally different from the
equilibrium one. We expect this feature to persist for the
model of Fig. 15, i.e it is not an artefact of the continuous limit !

Other quantities can be computed in this model. Let us give some
examples:

\medskip

{\it calculation of separation of replicas}

\medskip
 
One can also compute the separation of two thermal
replicas which are allowed to split at time $t_w$.
This quantity was studied in \cite{barrat}.

 One has:
 
 \begin{eqnarray}
 && Q_2 (z,t_w,\tau) = 
 \int dy dx P(y,t|x,t_w) P(y+z,t|x,t_w) P(x,t_w|x_0=-\infty,0) \\
 && = e^z \int_0^{+\infty} du_1 \int_0^{+\infty} du u_1 G(u_1,\tau|u) G(u_1 e^z ,\tau|u)
 G(u,t_w|0)  \\
 && G(u,\tau|u')  = \frac{1}{\sqrt{4 \pi \tau}} ( \exp(- \frac{(u-u')^2}{4 \tau} ) + 
 \exp(- \frac{(u+u')^2}{4 \tau} ) )
 \end{eqnarray}
 
 and one finds:
 
 \begin{eqnarray}
 && Q_2 (z,t_w,\tau) = \frac{e^z}{\pi} ( \frac{\Gamma}{(e^z - m)^2 + \Gamma^2}
 + \frac{\Gamma}{(e^z + m)^2 + \Gamma^2} ) \\
 && \Gamma = \sqrt{\frac{\tau (2 t_w + \tau)}{(t_w + \tau)^2}}
 ~~~~ m = \frac{t_w}{t_w + \tau}
 \end{eqnarray}
 
with $\tau=t-t'$. Note that this distribution is symmetric under
$z \to -z$ because of the relation $\Gamma^2 + m^2 =1$.
For large $\tau \gg t_w$ one finds that
the distribution of $x=e^z$ goes to a fixed half Cauchy unit distribution
$\theta(x) 1/(\pi(1+x^2))$. Thus
for large time we find that the two replicas evolve within a 
 finite distance, but this distance is {\it larger} than the dynamical $q_{EA}$ which
 is zero for this problem.

 Note that when $\tau/t_w \ll 1$ one recovers exactly the
 previous $Q(z,t,t')$.
 
\medskip

{\it non trivial FDT violation ratio ?}

\medskip

It is interesting  to know if one can find finite
dimensional models with a non trivial FDT violation
ratio $X$ as in mean field.

Here one can also compute the response to an additional
 field, i.e the remanent magnetization decay.
The calculation is indicated in the Appendix
\ref{appendixe}.
Though it does appear that in some sense this model
has a non trivial FDT violation ratio $X$ similar to mean field
we were not able to exhibit it in a clear way. If one 
looks at the finite fraction of the packet which
has not touched the reflecting boundary, it has clearly
a non trivial $X$. But on any global quantity we have looked at
the boundary effects always introduce a cutoff which changes the expected
result. We do not know if this is a purely technical
limitation, and if the model can be improved to really 
exhibit a non trivial and properly defined $X$ or if
this is a more fundamental limitation. 
We still present
some of the calculations in the Appendix \ref{appendixe} for the brave 
who is encourage to improve on it.

\subsection{solution of more general deterministic model}

\subsubsection{barriers growing faster than linear}

One can study cases where $E_b(x) \sim |x|^b$ with $b>1$.
From the Section \ref{barriers}  we expect that aging should have simple
properties. For the ``half landscape'' model, using 
$u(x) \sim \exp(x^b)$ for $x>0$ and a left reflecting wall at $u=1$,
one finds an aging regime with no thermal fluctuation: 

 \begin{eqnarray}
&& z = (\ln (\sqrt{t'} v_1 + \sqrt{t-t'} v_2)^{1/b} 
- (\ln(\sqrt{t'} v_1))^{1/b} \\
&& z = (\ln \sqrt{t})^{1/b} - (\ln \sqrt{t'})^{1/b} 
 \end{eqnarray}

where $v_1$ and $v_2$ are uncorrelated normalized gaussian variables. 
We have performed an expansion, in the regime of interest $t \gg t'$
e.g:

 \begin{eqnarray}
(\ln(\sqrt{t'} v_1))^{1/b} \sim (\ln \sqrt{t'})^{1/b}
+ \frac{\ln(v_1)}{(\ln \sqrt{t'})^{1 - 1/b}}
 \end{eqnarray}

and dropped the contributions of the noise parts $v_1$ and $v_2$ which
vanish in the limit of large $t$ and $t'$ since $b>1$.

Thus aging becomes thermally deterministic as:

 \begin{eqnarray}  \label{mf}
z = \ln \frac{h(t)}{h(t')}
 \end{eqnarray}

with $h(t) = \exp( C (\ln t)^{1/b} )$ and $C=(1/2)^{1/b}$. These
results are strikingly similar with the result (\ref{strikingly})
of the previous Section.
In fact the result (\ref{mf}) is also similar to what was found in mean field
in \cite{Cule}.

There is also a diffusion regime, which is identical to the end 
of the aging one (it merges smoothly into it). Indeed:

\begin{eqnarray}
  z = (\ln \sqrt{t})^{1/b} (1 - (\frac{\ln t'}{\ln t})^{1/b} 
 \end{eqnarray}

\subsubsection{barriers growing slower than linear (subaging)}

Similarly, in the subaging case where $E_b(x) \sim x^b$ with $b<1$
one finds:

\begin{eqnarray}
z = \ln (1 + \sqrt{\tau} v_2 + \sqrt{t'} v_1)^{1/b} - \ln (1 + \sqrt{t'} v_1)^{1/b}
 \end{eqnarray}

Performing expansions it gives:

\begin{eqnarray}
z =
\frac{1}{b} \frac{\sqrt{\tau} v_2}{ (1 + \sqrt{t'} v_1)  \ln (1 + \sqrt{t'} v_1)^{1/b-1} }
\end{eqnarray}

and thus

\begin{eqnarray}
z = \frac{1}{b} \frac{\sqrt{\tau}}{\sqrt{t'} \ln \sqrt{t'}} v
\end{eqnarray}

where $v$ is again a variable with a Cauchy distribution
(we have not determined the exact form using the wall).
This is compatible with a Lorentzian aging packet but with:

\begin{eqnarray}
\Gamma = \frac{1}{b} \frac{\sqrt{\tau}}{\sqrt{t'} \ln \sqrt{t'}}
\end{eqnarray}

consistent with:

\begin{eqnarray}
\Gamma = f [ \frac{h(t)}{h(t')} ] ~~~ h(t) = \exp( C (\ln t)^{1/b} )
\end{eqnarray}

with $a<1$, again performing an expansion in small $\tau/t_w$ and with
$\beta=1/2$.

 \section{conclusion}

In this paper we have investigated two time quantities
in several one dimensional diffusion models with
random and non random environments. These quantities
are functions of the waiting
time $t'=t_w$ after the initial localized condition at $t=0$
and a later time $t$. The (averaged) distribution 
$Q(z,t,t')$ of relative displacements $z=x(t)-x(t')$ between $t$ and $t'$
was studied.
Part of this study was numerical (in Sinai's model) 
and we have reached times up to $10^{15}$. Our results 
showed that the times reached in a previous simulation
\cite{toyaging} were vastly insufficient. 
Our conclusions are different from the one of \cite{toyaging}.
Part of the study was analytical: we have computed two time quantities
for a directed model related to the biased 
Sinai model and we have introduced several new models which
can be studied analytically. Our main results are the following.

We have identified three generic regimes
for large times $t_w \to \infty$, $t \to \infty$:

(i) At small separations $\tau = t-t_w \ll t_w$ a quasi-equilibrium 
regime. Evidence for that regime was found in Sinai's model.
In that regime $\overline{Q(z,t_w + \tau,t_w)}$ reaches
a limit $\overline{Q(z,\tau)}$ for $t_w \to \infty$. We have argued,
and checked numerically, that this distribution has some 
peculiarities. For large $\tau$ it does admit a limit
$\overline{Q(z)}$ but this limit exhibits an algebraic tail
originating from rare configurations of the disorder. The moments
of the relative displacement $\overline{<z^n(\tau)>}$
with $n>1/2$ grow unboundedly with $\tau$. We have also 
proposed an expression for the distribution
$\overline{Q(z,\tau)}$ based on arguments on periodic media.

We have also concluded, from our simulations and from
physical arguments, that in Sinai's model usual equilibrium
theorems hold in this quasi equilibrium regime (TTI and FDT).
We have obtained a generalized expression of these theorems
to probability distributions such as $Q(z,t,t')$,
and shown that in this regime $Q(z,\tau)$ and the response
function $R(z,\tau)$ obey
and exact differential relation. That these theorems
should hold in that regime is confirmed by recently obtained rigorous bounds,
as we have discussed.
This unveils an interesting situation of a quasi equilibrium
regime with a lot of internal structure,
wide fluctuations, and internal logarithmic diffusion 
(moments growing with $\tau$) which calls for further
studies.

(ii) at large time separations, $L(t) \sim L(t_w)$ 
there is a diffusion regime. There the displacements 
scale as $x(t) \sim x(t_w) \sim L(t)$ and there are
scaling forms for the probability distributions.
In the model of Section (\ref{solvable}) we have obtained this regime 
analytically.

(ii) finally there is an intermediate aging regime.
One should first look at this regime in the 
probability of staying in a finite neighborhood $z$ of the same point
between $t$ and $t'$ which is generically of the form 

\beq \label{agingf}  
Q(z,t,t')= F[z,  \frac{h(t)}{h(t')} ]
\eeq

In Sinai's model with a bias (and in the directed model
with algebraic distribution of waiting times)
one has aging with $h(t)=t$.
If the waiting times are even more widely distributed,
we find (\ref{agingf}) (for $z=0$) with a large class of functions $h(t) \gg t$.
Similarly we also find this behaviour in a solvable model in (\ref{solvable})
where a large class of functions $h(t)$ (including subaging $h(t) \ll t$) can be
obtained.

In the symmetric Sinai model (without a bias) we have found strong
numerical evidence for the aging behaviour (\ref{agingf})
with $h(t) \sim \ln t$ (for small finite $z$). In that model
we have even found a more striking result:

\beq \label{agingf2}  
Q(z,t,t')= Q_0(z) f[\frac{h(t)}{h(t')} ]
\eeq

i.e a decoupled form for the aging regime. This suggests
an interpretation of the aging regime in Sinai's
model as well equilibrated well which get emptied on
aging time scales. This picture should be checked further. 

Another consequence
of our result for Sinai's model is that, since
the aging regime must be compatible with the diffusion one
and we have found that $h(t) \sim L(t) \sim \ln t$,
there must be a {\it singularity} in the two time diffusion front at $z=0$ to allow for
a non trivial aging regime.

We have given a general explanation of these regimes
using scaling arguments on the next highest barrier encountered
by the particle. This allows to understand the aging form
in Sinai's model. It also strongly suggests that the aging
in Sinai's model could be studied analytically by only computing
the distribution of next highest barrier (a purely geometrical
feature of the energy landscape). These considerations also
lead to define a class of models for which
the distribution of these barriers can be computed exactly,
and allows for predictions of the aging forms (\ref{agingf}).

Though we did get a consistent picture of aging in Sinai's
model, we cannot rule out completely other regimes. 
For instance, we have not explored in more details the behaviour of the
moments of the displacement in Sinai's model. As in 
Section \ref{directed} one could say that at the very beginning of 
the aging regime (i.e $\ln t/\ln t_w \sim 1+\epsilon$ fixed,
a small fraction of particles have escaped
from their well and have experienced Sinai's diffusion to another
well. One then gets:

\beqa
\overline{  | <x(t) - x(t_w)>|^n } 
\sim (\frac{\ln t}{\ln t_w} -1)^\beta (\ln(t-t_w))^{2 n}
\sim (\ln t_w)^{2n -\beta} (\ln t - \ln t_w)^\beta
\eeqa

Thus by the same mechanism as
in Section \ref{directed} the various moments may have some
different aging behaviours. We have not attempted to obtain a precise estimate for $\beta$
but a rough estimate from our numerical simulations 
(Fig. 13 ) is consistent with $\beta =1$. If this is the case the
moment $n=1/2$ may have an aging behaviour as $t/t_w$.
The general issue of the matching bewteen the three regimes
defined here deserves to be investigated further.

Another open problem is the behaviour of the 
response in these regimes. It is important
to determine how to define properly, and study beyond mean field
the way the equilibrium theorems are violated. 
Since, as we have shown, 
sample to sample fluctuations play a strong role and one should
focus on distributions, one needs extensions of the mean field
ideas. As a first step we have given analytical expressions and
definitions of quantities adapted to low dimension and which 
measure these violations. A detailed numerical and further
analytical investigation of these quantities is deferred to
the future.

To summarize, we have found that some of the concepts defined in mean field
are still useful in low dimensional models, though they have to be
seriously adapted. We hope that this study will also help understand 
dynamical behaviour in low dimensional but more complex 
systems such as domain wall motion with disorder and coarsening in 
random spin systems, where ultra slow anomalous {\it diffusion processes} 
are expected to play 
a crucial role. We expect that the various regimes defined here should be present in these
systems as well.

We thank A. Barrat, A. Georges, L. Cugliandolo
and J. Kurchan for useful discussions.

 \appendix

 \section{Discrete version of Sinai model and diagonalization}
\label{appendixa}

In this Appendix we describe in detail the observables for
discrete hopping models. We also describe the numerical method
used in the paper. We establish some FDT relations and other useful
exact relations for discrete models.

Let us consider the Fokker Planck operator $H_{FP}$ defined in
(\ref{fplanck}). It can be written as:

\beqa
(H_{FP})_{n,m} P_m = \frac{d P_n}{dt} = - ( J_{n+1,n} - J_{n,n-1} )
\eeqa

where the current flowing from site $n-1$ to $n$ is by definition:

\beqa   \label{current1}
J_{n,n-1} = e^{\phi_n} P_{n-1} - e^{- \phi_n} P_n 
\eeqa

We need to compute the Green function $P(n,t|n_0,t_0)$
 ($t \geq t_0$) which 
 is defined as the solution of 
 (\ref{fokkerplanckd}) with initial 
 condition $P(n,t_0|n_0,t_0)= \delta_{n n_0}$. 
 Following Ref. \cite{bouchaud_1d} it is useful to
 map the FP equation onto a Schr\"{o}dinger equation corresponding
 to a symmetric matrix. One has:
 
 \beq
 P(n,t|n_0,t_0) = e^{- \frac{1}{2}(U_n - U_{n_0})} \sum_\alpha 
 \psi^\alpha_n \psi^\alpha_{n_0} e^{- E_\alpha t}
 \label{greenf}
 \eeq
 
 where the $\psi^\alpha$ are the eigenstates of the 
 Schr\"{o}dinger operator:
 
 \beq
 (H_s)_{n,m} \psi^\alpha_m = -
 ( \psi^\alpha_{n+1} + \psi^\alpha_{n-1} - 2 \psi^\alpha_{n} )
 + V_n \psi^\alpha_{n} = E_\alpha \psi^\alpha_{n}
 \label{schrod}
 \eeq
 
 in the potential:
 
 \beq
 V_n = e^{\phi_{n+1}} + e^{-\phi_{n}} - 2
 \eeq
 
 As discussed in Ref. \cite{bouchaud_1d} the 
 random operator $H_s$, which is
 a version of supersymmetric quantum mechanics,
 is quite peculiar :
 all states are localized but the spectrum of
 $H_s$ is positive, without the Lifschitz tails
 usually associated to random one-dimensional
 potentials. An eigenfunction corresponding to 
 the energy level $E_0 = 0$ is always 
 exactly known, i.e $\psi^0_{n}=Z e^{-U_n/2}$. Whether or
 not this is the actual ground state depends on whether
 $\psi^0_{n}$ is normalizable, i.e on the boundary conditions.
 This is also related to the breaking of supersymmetry.

 The Schr\"{o}dinger operator is a tridiagonal symmetric
 matrix and is easily diagonalized for large size $L$.
 Let us first consider the problem with $L+1$ sites $k=0,L$ 
 with reflexive boundaries. The same change of function
 $P_n = e^{U_n/2} \psi_n$ is used and formula (\ref{greenf})
 holds.
 
 \beqa
 (H_{FP})_{ij} &=&
  \delta_{i+1,j} e^{-\phi_j} + \delta_{i-1,j} e^{\phi_{j+1}}
 - \delta_{i,j} ( e^{-\phi_i} + e^{\phi_{i+1}} ) \nn
 && (H_{FP})_{0j} = \delta_{1,j} e^{-\phi_1} - \delta_{0,j} e^{\phi_1} \nn
 (H_{s})_{ij} &=&
 - ( \delta_{i+1,j} + \delta_{i-1,j} )
 + \delta_{i,j} ( e^{-\phi_i} + e^{\phi_{i+1}} ) \nn
 && (H_{s})_{0j} = - \delta_{1,j} + \delta_{0,j} e^{\phi_1}
 \eeqa
 these boundary conditions simply amount to choose $\phi_{-1}=+\infty$ and
 $\phi_{L+1}=-\infty$ and restrict the problem to sites $k=0,..L$.

\medskip
 
{\it explicit expressions of quantities of interest and FDT theorems in
discrete version}

\medskip

The two time quantities of interest are correlation functions
of some operator $O(n,n')$:

\beqa
 O(t,t') &=& < O(x(t),x(t'))>
 = \sum_{n,n'} O(n,n') P(n,t|n',t') P(n',t'|n_0,t_0=0) \nn
 && = \sum_{\alpha,\beta} e^{- E_\alpha (t-t') - E_\beta t'}
\sum_n \sum_{n'} O(n,n') e^{-U_n/2} \psi^\alpha_n \psi^\alpha_{n'} \psi^\beta_{n'}
 e^{U_{n_0}/2} \psi^\beta_{n_0}
\eeqa

and response functions of some operator $R(n,n')$

\beqa
R(t,t') &=& \frac{ \delta< R(x(t),x(t')) >_f  }{ \delta f(t') }
\eeqa

defined by adding a short-duration pulse of 
 small uniform force of integrated strength $f$, i.e
 $\phi_n \to \phi_n + f/2$ at time $t'$.
Using the definition
 $P(n,t|n',t') = ( e^{H_{FP} (t-t')} )_{nn'}$ one has:

 \beqa
 R^{+}(t,t') 
 =  \frac{1}{2} \sum_{n,n',m'} R(n,n') P(n,t|n',t') B_{n'm'} P(m',t'|n_0,t_0=0)
 \eeqa

if the pulse is at $t'-\epsilon$, or 

 \beqa  \label{formular}
 R(t,t') 
 =  \frac{1}{2} \sum_{n,n',m} R(n,n') P(n,t|m,t') B_{m n'} P(n',t'|n_0,t_0=0)
 \eeqa

if the pulse is at $t'+\epsilon$ (Ito).

We have defined

\beq
 B_{mm'} = \sum_{n'} \frac{ ( \delta H_{FP} )_{mm'} }{ \delta \phi_{n'} }
 = \delta_{m,m'} (e^{-\phi_m} - e^{\phi_{m+1} }) + 
 \delta_{m-1,m'} e^{\phi_m} - \delta_{m+1,m'} e^{-\phi_{m+1} }
 \eeq

We will now establish (i) the FDT relations valid when equilibrium is
attained (ii) some exact relations always valid.
These equations will relate the correlation functions of the operator
$O(n,n')$ with response function of the operator $R(n,n')$. There must
be a relation between $R$ and $O$ for these to hold, which is:

\beqa  \label{relation3}
\frac{1}{2} ( R(n,m) + R(n,m-1)) = O(n,m) - O(n,m-1)
\eeqa

This relation generalizes the usual relation $R(x,x')=\partial_{x'} O(x,x')$
(see next Section) valid for continuous systems.

We start with the following identity:

 \beq
 \frac{ \partial O(t,t')}{\partial t'} = 
 \sum_{n,n',m,m'} O(n,n') P(n,t|m,t') H_{m m'} ( \delta_{m n'} - \delta_{m' n'} )
P(m',t'|n_0,t_0=0)
 \eeq

Note the simplification:

 \beq
H_{m m'} ( \delta_{m n'} - \delta_{m' n'} ) = (\delta_{m+1,m'} e^{-\phi_{m'}}
+ \delta_{m-1,m'} e^{\phi_{m'+1}} )(\delta_{m n'} - \delta_{m' n'} )
\eeq

Now the following exact relation can be established:

 \beqa
\sum_{n',m'} O(n,n') H_{m m'} ( \delta_{m n'} - \delta_{m' n'} ) P_{m'}
= \frac{1}{2} \sum_{n'} R(n,n') B_{m n'} P_{n'}  + \frac{1}{2} R(n,m) (J_{m+1,m} + J_{m,m-1} )
\eeqa

for any set of $P_{m}$,
provided the above relation (\ref{relation3}) holds between the
operators $O$ and $R$ ($J$ being defined as in (\ref{current1})).
It yields to:

 \beqa
 \frac{ \partial O(t,t')}{\partial t'} = R(t,t') 
+ 
\frac{1}{2} \sum_{n,m} R(n,m) P(n,t|m,t') (J_{m+1,m}(t') + J_{m,m-1}(t'))
\eeqa

where $J_{m,m-1}(t')=e^{\phi_m} P(m-1, t'|n_0,t_0=0) - e^{- \phi_m} P(m, t'|n_0,t_0=0)$
and we are using the Ito response.

When equilibrium is attained, i.e either 
in the limit $t' \to + \infty$ {\it before} $L \to + \infty $
or, if there is an FDT regime in the problem (see text, this 
usually entails averaging all these correlations over
disorder), then one can set the current to zero, $J=0$. Then the 
the fluctuation-dissipation relation holds:

\beqa
 \frac{ \partial O(t,t')}{\partial t'} = R(t,t')
\eeqa

\medskip

{\it applications}

\medskip

- The choice $O(m,n)=n m$ and $R(n,m)=n$ is consistent
with (\ref{relation3}). It gives:

 \beqa
 \frac{ \partial <x(t) x(t')>}{\partial t'} = 
\frac{ \delta< x(t) >_f  }{ \delta f(t') }
\eeqa

- The choice $O(m,n)=(n-m)^2$ and $R(n,m)=-2(n-m)$ is consistent
with (\ref{relation3}). It gives:

 \beqa
 \frac{ \partial <(x(t)-x(t'))^2>}{\partial t'} = -2
\frac{ \delta< (x(t) - x(t') >_f  }{ \delta f(t') }
\eeqa

- The general choice:

\beqa
O(n,m) = \frac{1}{2} ( \tilde{O}(n,m) + \tilde{O}(n,m+1) )
~~~
R(n,m) = \tilde{O}(n,m+1) - \tilde{O}(n,m)
\eeqa

satisfies the condition (\ref{relation3}).

Thus one can choose $\tilde{O}(n,m) = \delta_{z-(n-m)}$
and obtain the exact relation:

\beqa
&& \partial_{t'} \frac{1}{2} ( Q(z,t,t') + Q(z+1,t,t') )
=
R(z+1,t,t') - R(z,t,t')  \\
&& + 
\frac{1}{2} \sum_{m} (P(z+1+m,t|m t') - P(z+m,t|m t'))
(J_{m+1,m}(t') + J_{m,m-1}(t'))
\eeqa

This is the discrete equivalent of the continuous
relation derived in the next Section (\ref{exactz}).

Finally, one can wonder what happens when a bias is applied on a
finite size periodic ring. There a {\it stationary} distribution 
with a fixed current is reached at large $t'$. There is
an extension of the FDT theorem. Indeed one has:

 \beqa
 \frac{ \partial O(t,t')}{\partial t'} = R(t,t') 
+ J \sum_{n,m} R(n,m) P(n,t|m,t')
\eeqa

Note that $J$ is simply related to the velocity 
(see e.g \cite{ledou_vino}).

 \section{FDT and useful exact relations for probability distributions}
\label{appendixb}
 
 In this Appendix we derive a generalization of the FDT theorem
 on the probability distribution. It can then be used as a
 generating functional to obtain a hierarchy of FDT relations
 on all moments of the type $<x(t)^n x(t')^m>$. 

 We are interested in the joint probability that the particle
 is in $x$ at $t'$ and then in $x$ at $t$.
 
 \begin{eqnarray}
 \hat{P}(xt,x't'|x_0t_0) = P(x,t|x',t') P(x',t'|x_0,0)
 \end{eqnarray}
 
 Let us recall the forward and backward FP equations:
 
 \begin{eqnarray}
 && \partial_t P(xt|x't') =
 T \partial_x D(x) \partial_x P(x t|x't') - \partial_x D(x) F(x) P(x t|x't') \\
 && \partial_{t'} P(xt|x't') =
 -T \partial_{x'} D(x') \partial_{x'} P(x t|x't') - D(x') F(x') \partial_{x'}  P(x t|x't') 
 \end{eqnarray}
 
 We now derive the differential equation for the joint probability $\hat{P}$:
 
 \beq
 \partial_{t'} \hat{P} = (\partial_{t'} P(x,t|x',t')) P(x',t'|0,0)
 + P(x,t|x',t') (\partial_{t'} P(x',t'|0,0))
 \eeq
 \beq
 \partial_{t'} \hat{P} = - \int dy dy' P(x,t|y,t') ( [H_{FP}, \delta_{x'}])_{yy'} 
  P(y',t'|0,0)
 \eeq
 
 with:
 \beq
 [H_{FP}, \delta_{x'}] = [ T \partial D \partial - \partial D
 F, \delta_{x'}] = T \partial D [\partial,\delta_{x'}] +
 [\partial,\delta_{x'}] (T D \partial - D F)
 \eeq
 
The last term is the current $J=- (T D \partial - D F)P$. In 
the FDT regime (for large $t'$) the current is expected to vanish and
we are left with the following equation for $\hat{P}$:
 
 \begin{eqnarray}
 \partial_{t'} \hat{P} =  - \int dy dy' P(x,t|y,t') (  T \partial D
 [\partial,\delta_{x'}])_{yy'} P(y',t'|0,0)\\
 \partial_{t'} \hat{P} =  \partial_{x'} \int dy dy' P(x,t|y,t') (T \partial D
 \delta_{x'})_{yy'} P(y',t'|0,0)\\
 \partial_{t'} \hat{P} = - T \partial_{x'} [ ( \partial_{x'} P(x,t|x',t')) D(x')
 P(x',t'|0,0) ]
 \end{eqnarray}
 
 finally we obtain:
 \beq  \label{exact2}
 \partial_{t'} \hat{P} = - T \partial_{x'}^2 (D(x') \hat{P}) + T \partial_{x'} (
 P(x,t|x',t') \partial_{x'} ( D(x') P(x',t'|0,0)) )
 \eeq
 
 and since $\frac{\delta H_{FP}}{\delta h(t')} =  - \partial_{x'} D(x')$
 
 \beq
 \partial_{t'} \hat{P} = - T \partial_{x'}^2 (D(x') \hat{P}) - T \partial_{x'}
 \frac{\delta \hat{P}^{+}_h}{\delta h(t')}
 \eeq
 
where we define $\hat{P}^{+}_h(xt|x't'|x_00) $ the joint probability
when a field pulse has been applied at $t'-\epsilon$.
 This is the equation which relates exactly the joint probability
 distribution $\hat{P}(xt,x't'|x_0t_0)$ to the response distribution
 in the quasi-equilibrium FDT regime. It was obtained by setting the
 current at time $t'$ to zero.

 A similar equation can be derived for the field applied at time
$t'+\epsilon$. From (\ref{exact2}) one has also that:

 \beq
 \partial_{t'} \hat{P} = - T \partial_{x'}
 \frac{\delta \hat{P}_h}{\delta h(t')}
 \eeq

where 
$\hat{P}_h(xt|x't'|x_0 0) $ the joint probability
when a field pulse has been applied at $t'+\epsilon$.
This corresponds to Ito's prescription for the response
functions since 
$\delta \hat{P}_h(xt'|x't'|x_0 0)/\delta h(t')=0$.

 From this it is immediate to derive a similar FDT equation
 for the probability function $Q(z,t,t')=
 \int dx dx' \delta(z-x+x') \hat{P}(xt,x't'|x_0t_0)$.
 Multiplying the above equation by the delta function,
 integrating with respect to x and x', and integrating
 by parts one gets:
 
 \beqn  \label{current2}
 \partial_{t'} Q(z,t,t') = - T \partial_z^2   Q(z,t,t')
 + T \partial_z \frac{\delta Q^{+}_h (z,t,t')}{\delta h(t')} 
 + T \partial_z \int dx_1 dx_2 \delta(z-x_1+x_2) P(x_1,t|x_2,t')  J(x_2)
 \eeqn
 
 in the the long time regime $t' \rightarrow \infty$ 
 the current vanishes and we are left with the FDT regime
 of the equation:
 
 \beq
 \partial_{t'} Q(z,t,t') = - T \partial_z^2   Q(z,t,t')
 + T \partial_z \frac{\delta Q^{+}_h (z,t,t')}{\delta h(t')}
 \eeq
 
Similarly with the Ito prescription:

 \beq  \label{relative}
 \partial_{t'} Q(z,t,t') = 
  T \partial_z \frac{\delta Q_h (z,t,t')}{\delta h(t')}
 \eeq

 We can check that this equation is more general
 than the conventional FDT theorem, and indeed gives
 back the usual FDT result. Defining:
 \beq
 B(t,t')_{FDT} = \int dz z^2 Q_{FDT} (z,t,t')
 \eeq
 
 and inserting it in the generalized equation gives:
 
 \beq 
 \partial_{t'} B(t,t') = -  T \int dz z^2 \partial_z^2   Q(z,t,t')
 + T \int dz z^2 \partial_z \frac{\delta Q^{+}_h (z,t,t')}{\delta h(t')}
 \eeq
 after integration by part this gives simply:
 
 \beq 
 \partial_{t'} B(t,t') = -2 T  -2 T 
 \frac{\delta}{\delta h(t')}  \int dz z Q^{+}_h (z,t,t')
 = - 2 T ( 1 + R^{+}(t,t') - R^{+}(t',t'-\epsilon) ) =
 - 2 T R(t,t')
 \eeq
 
With the Ito prescription one has simply:

\beq 
 \partial_{t'} B(t,t') =  -2 T 
 \frac{\delta}{\delta h(t')}  \int dz z Q_h (z,t,t')
 = - 2 T R(t,t')
 \eeq

using $R(t',t')=R(t',t'+\epsilon)=0$.

 A motivation is to find a generalized form to this equation
 which would be valid in the aging regime as well.

\medskip

{\it exact relations and FDT violation ratios}

\medskip

It is useful also to give the exact relations (always valid)
for averages of operators. They allow to obtain explicit 
the FDT violation ratios.

We put back the current term that we have neglected. We obtain
then instead of (\ref{exact2}) the (still) exact relation:

 \beq
 \partial_{t'} \hat{P} + T \partial_{x'}^2 (D(x') \hat{P}) - T \partial_{x'} (
 P(x,t|y,t') \partial_{x'} D(x') P(x',t'|0,0)) = 
\int dy dy' P(x,t|y,t') [\partial,\delta_{x'}]_{yy'} J(y't'|0 0)
\eeq

with $J(y't'|0 0) = - D(y') (T \partial_{y'} - F(y'))P(y't'|0 0)$.
We will use the Ito response here.

Let us study a general observable $O(x,x')$. Multiplying
the above equation, integrating over $x$ and $x'$, and integrating by parts
(assuming no contributions from boundaries) one has an exact relation 
which relates the averages of
of $O(x,x')$ and of $O'(x,x')= \partial_{x'} O(x,x')$. Defining:

\beq \label{viola}
V_0(t,t') = \partial_{t'} <O(x(t),x(t'))> -T \frac{ \delta <O'(x,x')>_h }{\delta h(t')}
\equiv (1 - X_O(t,t')) \partial_{t'} <O>(t,t')
\eeq

the relation reads:

\beq  \label{current}
V_0(t,t') =  \int dx dx'  P(x,t|x',t') O'(x,x') J(x't'|0 0)
\eeq

We have defined above the FDT violation ratio $X_0$ associated with the
operator $O$. 
If the other response was used there would be in addition
a term  $- T < D(x(t')) O_2(x(t),x(t')) >$ in the above equation
with $O_2(x,x') = \partial^2_{x'} O(x,x')$.
Again for the Ito response the second derivative term is 
absent.

We will also give an exact relation for the {\it relative displacements}.
Let us consider an operator $O(z)$. Then one has:

\beqa  \label{exactz}
\partial_{t'} <O(z)> + T \frac{\delta < O'(z) >_h}{\delta h(t')}
= \int dx_1 dx_2 O'(x_1 - x_2) P(x_1 t|x_2 t') J(x_2 t' |00) 
\eeqa

And thus one can define also a generalized FDT ratio:

\beqa \label{tildex}
\tilde{X}_O(t,t') = 1 - \frac{ \int dx_1 dx_2 O'(x_1 - x_2) P(x_1 t|x_2 t') J(x_2 t' |00) }{
\partial_{t'} <O(z)> }
\eeqa

\section{Bounds} \label{appendixc}

Here we illustrate the bounds recently proposed by CDK \cite{Cudeku}. We use the
framework of the generalized FDT relation of the preceding section and 
our derivation is thus technically slightly different, though identical in
spirit to \cite{Cudeku}. We work directly with the FP equation and a
space dependent diffusion coefficient.

The nice observation of CDK is that the current which appear in
(\ref{current}) also appears in the $H$ theorem which states that the free
energy:

 \beq
H(t') = \int dx' P(x't'|0 0) (T \ln P(x',t'|0 0) - U(x'))
\eeq

is always decreasing with:

\beq
\frac{d H(t')}{dt'}  = - \int dx' \frac{(J(x't'|0 0))^2}{D(x') P(x',t'|0 0)}
\eeq

The CKS bounding amounts to bound $V_0(t,t')$ defined in 
(\ref{viola}) by:

\beqa
&& | V_0(t,t') | \leq 
|\frac{d H(t')}{d t'}|^{1/2} 
\int dx dx' P(x,t|x',t')  O'(x,x')^2 D(x') P(x',t'|00) \\
&& = |\frac{d H(t')}{d t'}|^{1/2}  
<  (O'(x(t),x(t'))^2 D(x(t')) >^{1/2}
\eeqa

using the Cauchy-Schwartz CS inequality and $\int dx P(x,t|x't') =1$

Note also that disorder averages can be bounded similarly 
by applying the CS inequality at the same time to the
integrals over $x,x'$ and configurations.

One then gets:

\beqa
&& | \overline{V_0(t,t')} | \leq |\frac{d \overline{H}(t')}{d t'}|^{1/2}  
| \overline{ <  O'(x(t),x(t'))^2 D(x(t')) >} |^{1/2}
\eeqa

Or, as pointd out by CDK, in an integrated version:

\beqa
&& | \int_{t'}^{t} ds \overline{V_0(t,s)} | \leq \int_{t'}^{t} ds
( |\frac{d \overline{H}(s)}{d s}|^{1/2}  
| \overline{ <  O'(x(t),x(s))^2 D(x(s)) >} |^{1/2} )
\eeqa

If we choose $O(x,x')=x x'$ and $D(x)=1$ one gets:

\beqa
| \overline{<x^2(t)>} - \overline{<x(t) x(t')>}
- T \int_{t'}^{t} R(t,t')| \leq (\overline{<x^2(t)>})^{1/2} \int_{t'}^{t} ds
( |\frac{d \overline{H}(s)}{d s}|^{1/2} 
\eeqa

One can also derive bounds using (\ref{exactz})
for the {\it relative displacements}.
Using CS,  (\ref{exactz}) leads to the bound:

\beqa
|\partial_{t'} <O(z)> + T \frac{< \delta O'(z) >_h}{\delta h(t')}| 
\leq |<(O'(x(t) - x(t'))^2 >|^{1/2} |\frac{d \overline{H}(t')}{d t'}|^{1/2}
\eeqa

Which can be rewritten as:

\beqa
| \partial_{t'} \int dz O(z) Q(z,t,t')
+ T \frac{\delta}{\delta h(t')} \int dz O'(z) \frac{\delta Q_h(z,t,t')}{\delta h(t')}|
\leq |\int dz O'^2(z) Q(z,t,t')|^{1/2} |\frac{d \overline{H}(t')}{d t'}|^{1/2}
\eeqa

This yields in particular:

\beqa
| \partial_{t'} B(t,t') + 2 T R(t,t') | \leq B(t,t')^{1/2}
|\frac{d \overline{H}(t')}{d t'}|^{1/2}
\eeqa

Or its integrated version:

\beqa
| - B(t,t') + 2 T \int_{t'}^{t} R(t,t') | \leq \int_{t'}^{t} ds 
B(t,s)^{1/2} |\frac{d \overline{H}(s)}{d s}|^{1/2}
\eeqa

More generally the bound can be used to constrain the $\tilde{X}_{O}$
defined in (\ref{tildex}):

\beqa
|1 - \tilde{X}_{O}(t,t')| \leq 
\frac{ <O'^2>(t,t') |\frac{d \overline{H}(t')}{d t'}|^{1/2}
 }{| \partial_{t'} <O>(t,t') |} 
\eeqa

 \section{mappings of several models} \label{appendixd}
 
 The method of change of variables allows to relate exactly
 members of a class of landscape. Let two landscapes and
their corresponding Green's functions be:

 \begin{eqnarray}
 (E_b(x) , U(x) ) & \leftrightarrow & P(x, t |x_0, 0 ) \\
 (E'_b(x) , U'(x) ) & \leftrightarrow & P'(x,t |x_0, 0)
 \end{eqnarray}
 
 If there exist a function $y(x)$ such that:
 
 \begin{eqnarray}
 E'_b(x) = E_b(y(x)) + 2 \ln \frac{dy(x)}{dx} ~~~
 U'(x) = U(y(x)) - \ln \frac{dy(x)}{dx}
 \end{eqnarray}
 
 then the two Green functions are related through:
 
 \begin{eqnarray}
 P'(x, t |x_0, 0) = \frac{dy(x)}{dx} P(y(x), t| y(x_0) , 0)
 \end{eqnarray}

 In particular one can map:
 
 \begin{eqnarray}
 (0, U(x) ) & \leftrightarrow & (2 U(x(u)) , 0)
 \end{eqnarray}
 
 and 
 
 \begin{eqnarray}
 (E_b(x) , 0 ) & \leftrightarrow & (0 ,  \frac{1}{2} E_b(x(u)) )
 \end{eqnarray}
 
 It is interesting in general because the new functions
 $U(x(u))$ or $E_b(x(u))$ are usually better behaved at large $u$.

 Two general scenarios exist, confining potentials,
 unconfining ones. Let us take rapidly growing landscapes ($E_b(x)$, $U(x)$)
 as in last section and map them onto (0, $U'(u)$). Then
 typically one has $U'(u) \sim b \ln u$ at large $u$. The case
 $b 0$ is confining and corresponds to the case where
 barriers grow faster than valleys $E_b(x)  - 2 U(x)$. The case
 $b 0$ is fast diffusion and corresponds to the case where
 barriers grow slower than valleys $E_b(x) <  - 2 U(x)$.
 
 Finally note that time dependent mappings could also be
built using the following propagator:
 
 \begin{eqnarray}
 Q(u,t) = \frac{1}{\sqrt{4 \pi e^{t-t_0} - 1} }
 e^{- \frac{ (u - u_0 e^{(t-t_0)/2} )^2 }{ 4 (e^{t-t_0} - 1) } }
 \end{eqnarray}

 which satisfies:
 
 \begin{eqnarray}
 \partial_t Q(u,t) =
 \partial_u \partial_u Q(u,t) - \frac{1}{2} \partial_u u Q(u,t)
 \end{eqnarray}

 \section{response in the aging model} \label{appendixe}
 
 We start from the exact equation obeyed by
 the Brownian diffusion propagator:
 
 \begin{eqnarray}
 R(ut|u't') = \frac{1}{\sqrt{4 \pi T (t-t')}} e^{- \frac{(u-u')^2}{4 T (t-t')}}
 \end{eqnarray}
 
 The joint propagator 
 
 \begin{eqnarray}
 \hat{R}(ut|u't'|u_0 t_0) = R(ut|u't') R(u't'|u_0 t_0)
 \end{eqnarray}
 
 satisfies the exact equation:
 
 \begin{eqnarray}
 \partial_{t'} \hat{R} = - T \partial_{u'}^2 \hat{R} + 2 T 
 \partial_{u'} ( R(ut|u't') \partial_{u'} R(u't'|u_0 t_0) ) \\
 \partial_{t'} \hat{R} = - 2 T 
 \partial_{u'} \frac{ \delta \hat{R}_h }{\delta h(t')} 
 \end{eqnarray}
 
 This is valid for the free (unbounded) brownian motion and
 yields for instance $R(t,t') = X \partial_{t'} C(t,t')$ with $X=1/(2 T)$.
 A bounded brownian motion would instead converge to 
 equilibrium with $X = 1/T$ and satisfy the
 FDT equation with $2 T$ replaced by $T$ in the last term.

 One can now use the change of variable $\Phi(x) d/dx = d/du$ 
 and $\hat{R}(ut|u't'|u_0 t_0) = \Phi(x) \Phi(x') \hat{P}(xt|x't'|x_0t_0)$
 and obtain for the model studied previously, the exact equation
 valid in the unbounded case:
 
 \begin{eqnarray}
 && \partial_{t'} \hat{P} = - T \partial_{x'}^2 ( \Phi(x')^2 \hat{P} )
 + T \partial_{x'} ( P(xt|x't') \partial_{x'} ( \Phi(x')^2 P(x't'|x_0 t_0) ) ) \\
 && + T \partial_{x'} ( P(xt|x't') \Phi(x') \partial_{x'} ( \Phi(x') P(x't'|x_0 t_0) ) )
 \end{eqnarray}
 
 Since the diffusion coefficient is $D(x') = \Phi(x')^2$ in this
 model, and the response is
 $\frac{\delta H_{FP}}{\delta h(t')} = - \partial_{x'} D(x')$
 the above equation can be rewritten in a form very similar - but
 not identical - to
 the above general FDT equation:
 
 \beq
 \partial_{t'} \hat{P} = - T \partial_{x'}
 \frac{\delta \hat{P}_h}{\delta h(t')} 
 -  \partial_{x'} ( P(xt|x't') J(x't') )
 \eeq
 
 The last term cannot be simplified further and involves the
 current $J(x't') = - T \Phi(x') \partial_{x'} ( \Phi(x') P(x't'|x_0 t_0) )$
 of the model. This equation is always valid for our model, even in the out of
 equilibrium regime.
 
 The idea is that in the non trivial aging regime
 all three terms of the above equation will be roughly of the same
 order in $t'$ and thus it will effectively lead to a non trivial
 FDT ratio $X(t,t')$.

 One can use e.g this equation to study the correlation
 $C(t,t') = < x(t) x(t')> $. Multiplying by $x x'$ and
 integrating over $x$ and $x'$ one obtains:

 \beq
 R(t,t') \equiv \frac{\delta < x(t) _h }{\delta h(t')}
 = X(t,t') \partial_{t'}  < x(t) x(t')  >
 \eeq
 
 where we have defined:
 
 \beq
 X(t,t') = \frac{1}{T} (1 - \frac{ 
 \int dx dx' x P(x t|x't') J(x't') }{\partial_{t'}  < x(t) x(t') > } )
 \eeq
 
One can also define:

 \begin{eqnarray}
 X_B(t,t') = - \frac{1}{2} X_C(t,t') (1 -
 \frac{ \partial_{t'} < x(t')^2  }{2 \partial_{t'} < x(t) x(t')} )^{-1}
 \end{eqnarray}

 Introducing:
 
 \begin{eqnarray}
 && A(t',\tau) =
 < x[\sqrt{t'} u_1 + \sqrt{\tau} u_2]
 u_1 \Phi[ x[\sqrt{t'} u_1] ]  \\
 && D(t',\tau) =
 < (u_1 - u_2 \sqrt{\frac{t'}{\tau}} )
 x[\sqrt{t'} u_1] \Phi[ x[\sqrt{t'} u_1 + \sqrt{\tau} u_2] ] 
 \end{eqnarray}
 
 One finds:
 
 \begin{eqnarray}
 && X_C = \frac{1}{T} \frac{D}{A+D} \\
 && X_B = - \frac{1}{2 T} \frac{D}{D+A-A_0}
 \end{eqnarray}
 where $A_0 = A(t',0)$.
 
 Thus one gets for instance:
 
 \begin{eqnarray}
  X_B \sim - \frac{x}{2 T} 
 \frac{ <\epsilon(u_1) \epsilon(u_1+ x u_2) \ln(1 + \sqrt{t'}|u_1+ x u_2|) 
 - < \ln(1 + \sqrt{t'}|u_1|)  }{
 < \frac{x u_1 - u_2}{u_1 + x u_2} \epsilon(u_1) \epsilon(u_1+ x u_2)
 \ln(1 + \sqrt{t'}|u_1+ x u_2|)  }
 \end{eqnarray}
 
 with $x = \sqrt{\tau}{t'}$. The divergences make the calculation 
depend strongly on the boundary conditions. As explained in the
text we have not pursued it further.

 \section{directed model calculations} \label{appendixf}

Defining:

 \beq
 \Phi(s) = \overline{ \frac{1}{s+W} }
 \eeq

We will use that:
 
 \beqa
 \overline{ \frac{W}{s+W} } &=& 1 - s \Phi(s) \nn
 \overline{ \frac{1}{(s_1+W)(s_2+W)} } &=&
 \frac{ \Phi(s_1) - \Phi(s_2) }{s_2 - s_1} \nn
 \overline{ \frac{W}{(s_1+W)(s_2+W)} }  &=&
 \frac{ s_2 \Phi(s_2) - s_1 \Phi(s_1) }{s_2 - s_1} 
 \eeqa
 
 Let us compute the averaged probability that the particle
 advances by $m$ between $t'$ and $t=t'+\tau$:
 
 \beq
 Q(m,\tau,t') = \langle \overline{ \delta( x(t) - x(t') - m )}
 \rangle = 
 \sum_{n \ge 0} \overline{ P(n+m,n,\tau) P(n,0,t') }
 \eeq
 
 The double LT, $Q(m,s_1,s_2) = \int_0^{\infty} \int_0^{\infty} 
 d\tau dt' e^{-s_1 \tau - s_2 t'} P(m,\tau,t')$ can be
 calculated:
 
 \beqa
 Q(m,s_1,s_2) &=& \sum_{n \ge 0} \overline{P(n+m,n,s_1) P(n,0,s_2)} =
 ( \delta_{m0} ( \overline{ \frac{1}{(s_1+W)(s_2+W)}} )\nn
 &&+ (1 - \delta_{m0} ) ( \overline{\frac{1}{s_1+W}} )
 ( \overline{ \frac{W}{(s_1+W)(s_2+W)} } )
 (\overline{ \frac{W}{s_1+W} })^{m-1} )
 \sum_{n=0}^{\infty} (\overline{\frac{W}{s_2+W}})^n
 \eeqa

It yields the result given in the text.

Let us now estimate the probability $P_{t_w}(W)$ that at time $t_w$ the
walker is on a site with a waiting time $W=1/\tau$.

Its Laplace transform with respect to $t_w$ (Laplace
variable $s_2$) is simply given by:

\beqa
\overline{ \sum_n \delta(W - W_n) \frac{1}{s_2 + W_n}  \prod_{k=n_0}^{n-1} 
 \frac{W_k}{s_2+W_k} }
\eeqa
 
This easily leads to:

\beqa
\frac{1}{s_2} P(W) \frac{1}{s_2 + W} \frac{1}{\Phi(s_2)}
\eeqa
 
By Laplace inversion this yields for the distribution of the waiting time 
$\tilde{\tau}=1/W$:

\beqa
P_{t_w}(\tilde{\tau}) d\tilde{\tau} =  \frac{\sin(\pi \mu)}{\pi} \frac{d\tilde{\tau}}{t_w} 
 (\frac{t_w}{\tilde{\tau}})^{1+\mu}
\int_0^{1} e^{- \frac{t_w}{\tilde{\tau}} (1- u) } \frac{u^{\mu-1}}{\Gamma[\mu]}
\eeqa

\end{document}